\documentclass[referee]{aa} 
\usepackage{graphicx}
\usepackage{amsmath,epsfig,rotating,amssymb}

\begin{document}
   \title{Influence of Alfv\'en waves on the Thermal Instability in the Interstellar Medium}

   \author{Patrick Hennebelle\inst{1} and Thierry Passot\inst{2} }

   \offprints{P. Hennebelle}

   \institute{Laboratoire de Radioastronomie Millim\'etrique, UMR 8112 du CNRS, 
    \'Ecole Normale Sup\'erieure et Observatoire de Paris, 24 rue Lhomond,
              75231, Paris C\'edex 05, France
              \email{patrick.hennebelle@ens.fr}
	      \and 
              CNRS, Observatoire de la C\^ote d'Azur, B.P.\ 4229, 06304, Nice
              Cedex 4, France\\
              \email{passot@obs-nice.fr}}

   \abstract{The effect of Alfv\'en waves on
the thermal instability of the Interstellar Medium (ISM) is
investigated both analytically and numerically. A
stability analysis of a finite amplitude circularly polarized Alfv\'en
wave propagating parallel to an ambient magnetic field in a
thermally unstable gas at thermal equilibrium is performed, leading
to a  dispersion relation which depends on 3 parameters, namely the square 
ratio of the sonic and Alfv\'en velocities ($\beta$), the
wave amplitude and the ratio between the wave temporal  
period and the cooling time.  Depending on the values of these 3 parameters,
the Alfv\'en waves can stabilize the large-scale perturbations, destabilize
those whose wavelength is a few times the Alfv\'en wavelength  
$\lambda _{AW}$  or leave the growth rate of the short scales unchanged.
In order to investigate the non-linear regime, two different 
numerical experiments are performed in a slab geometry. The first
one deals with the development of an initial density perturbation in
a thermally unstable gas in the presence of  Alfv\'en waves. The second one
addresses the influence of those waves on the thermal transition
induced by a converging flow. The numerical results confirm the
trends inferred from the analytic calculations, i.e. the waves
prevent the  instability at scales larger than $\lambda _{AW}$
and trigger the growth of wavelengths close to $\lambda _{AW}$,  
therefore producing a very fragmented cold phase. The second
numerical experiments shows that i) the magnetic pressure  
prevents the merging of the CNM fragments therefore maintaining the
complex structure of the flow and organizing it 
into group of clouds ii) these groups  of CNM clouds have an
Alfv\'enic internal velocity dispersion  iii) strong density
fluctuations  ($\simeq 10 \rho_{\rm cnm}$)  triggered by  magnetic
compression occur. We note that  during this event there is no stiff
variation of the longitudinal velocity field. This is at variance
from the hydrodynamical case  for which the clouds are uniform and
do not contain significant internal motions except after cloud
collisions. In this situation a strong density fluctuation
occurs accompanied by a stationary velocity gradient through the cloud. 
}
\titlerunning{Influence of Alfv\'en waves on the thermal instability}
\authorrunning{Hennebelle \& Passot}
\maketitle

\section{Introduction}\label{intro}

It is now well established that the atomic interstellar medium is a
thermally bistable gas  which at a pressure of about 4000 K
cm$^{-3}$ can be in two different phases namely the Warm Neutral
Medium (WNM) and the Cold Neutral Medium (CNM) roughly in pressure equilibrium
(Kulkarni \& Heiles 1987, Field et al. 1969,  
Wolfire et al. 1995, 2003). It is also believed that the atomic gas
is strongly magnetized with magnetic intensity  
around 5$\mu$G (Troland \& Heiles 1986, Heiles 1987, Heiles \&
Troland 2005). Although the structure of the magnetic field 
is poorly known, some observational evidences seem to indicate that
the fluctuating part is at least comparable to the uniform one
suggesting that magnetized waves may be of great importance
for the dynamics of the ISM. Another important observational fact is
the absence of correlation between the  
magnetic intensity and the density in the interstellar atomic gas.  
Various works have investigated the effect of MHD waves on the 
dynamics of a polytropic or nearly polytropic (non thermally unstable)
magnetized gas (Dewar 1970, Goldstein 1978, McKee \& Zweibel 1993, Passot et al. 1995, 
Falle \& Hartquist 2002, Passot \& V\'azquez-Semadeni 2003).
 
It is thus of great interest to investigate the simultaneous
role of magnetic fields and of the bistable nature of the flow on
the physics of the ISM. Field (1965) considers the effect of a transverse
magnetic field on the thermal instability and generalizes the isobaric criterion. 
In the context of cooling flows, Loewenstein (1990) studies the
thermal instability in the presence of a static field. 
Hennebelle \& P\'erault (2000) investigates the role of an initially
uniform magnetic field, analytically and numerically, when the
thermal condensation is dynamically triggered. They propose a
mechanism based on magnetic tension to explain the thermal collapse
in a magnetized flow and  argue that  the magnetic intensity in the
WNM and in the CNM should not be very different. Piontek \& Ostriker
(2004, 2005) study the magneto-rotational instability  in a thermally unstable gas.

Here we investigate the effect of a finite amplitude circularly polarized Alfv\'en wave 
on the thermal instability, both analytically and
numerically. These Alfv\'en waves, which are exact solutions of the MHD
equations, are non dissipative and are thus very likely to be
present in a magnetized gas such as the ISM. Similar waves, although
at a smaller scale and in a regime affected by dispersive effects,
exist in the solar wind upstream of the earth bow
shock (Spangler et al. 1988). In-situ satellite observations have
clearly identified circularly polarized quasi-monochromatic Alfv\'en wave 
packets probably generated by reflected protons. 

In Sect. 2 we present a stability analysis of a
circularly polarized Alfv\'en wave  
propagating in a thermally unstable gas at thermal equilibrium and obtain a dispersion
relation which generalizes the relation obtained by Field (1965). We
then solve  this equation numerically for various regimes and
discuss the consequences for the ISM. 
In Sect. 3, we perform various numerical experiments to
confirm the analytic prediction and to investigate the  
non-linear regime. Section 4 concludes the paper.

\section{Analysis}\label{sec:num}

The equations governing, in the magnetohydrodynamic (MHD) limit,
the one-dimensional motion of a plasma permeated by a
uniform magnetic field ${\bf B_0}$ in a slab geometry read

\begin{eqnarray}
&&{\partial\rho\over\partial t} +{\partial(\rho u) \over\partial x}=0
\label{eq:mhd1d1}\\
&&{\partial u\over\partial t}+u{\partial u\over\partial x}
=-{1\over\rho} {\partial\over \partial x} \left ( P
+\frac{|b|^2}{8\pi} \right ) \label{eq:mhd1d2}\\
&&{\partial v\over\partial t}+u{\partial v\over\partial x} =
{B_x\over 4\pi\rho}{\partial b\over \partial x} \label{eq:mhd1d3}\\
&&{\partial b\over\partial t}+ {\partial\over \partial
x}(ub)=B_x{\partial v\over \partial x},
\label{eq:mhd1d4}\\
&&  (\frac{\partial T}{\partial t}+u\frac{\partial T}{\partial
  x})=-(\gamma-1)T\frac{\partial u}{\partial x} - \frac{1}{C_v}{\cal L} +
\frac{1}{\rho C_v} \partial _x (\kappa(T) \partial_x T )
\end{eqnarray}\par

\noindent where all fields only depend on the coordinate $x$ and
time $t$.

The ambient field ${\bf B_0}$ is assumed to be oriented in
the $x$ direction.
The velocity field ${\bf V}$ has a component  $u=V_x$ along the
$x$ coordinate and two transverse components combined in
the complex number $v=V_y+iV_z$. Similarly, we write the magnetic field 
as $b=B_y+iB_z$, the component $B_x=B_0$ remaining constant.
The mass density and thermal pressure are denoted by
$\rho$  and $P$ respectively and we assume a perfect gas law 
$\displaystyle{P=\frac{R \rho T}{\mu}}$ where
$\displaystyle{R=\frac{k_B}{m_H}}$ is the universal 
gas constant, $k_B$ the Boltzmann constant and $\mu$ the mean
molecular weight in units of the hydrogen mass $m_H$. Heating and
cooling processes are combined in a single net cooling 
function ${\cal L}$. The parameter $\gamma$ denotes the ratio of
the specific heats  at constant pressure $C_p$ and at constant
volume $C_v$. Note that $\displaystyle{C_v=\frac{R}{(\gamma-1)\mu}}$.
Thermal diffusivity  due to  neutrals is the dominant one. It is isotropic 
in spite of the presence of the magnetic field and 
equal to $\kappa(T)=5/3 C_v \eta(T)$, where 
$\eta(T)=5.7 \; 10^{-5} (T/1 K)^{1/2}$ g cm$^{-1}$ s$^{-1}$ (Lang 1974). 

Throughout most of Sect. 2, thermal diffusivity will be neglected for simplicity,
leading to assume a vanishingly small Field length.
Its effect will briefly be addressed in Sect. 2.5 and it is
explicitely taken into account in the numerical simulations of 
Sect. 3.

Two main non-dimensional numbers can be defined, the sonic Mach number
$M_s={V_0/c_s}$ ratio of a typical velocity $V_0$ with the constant
sound speed $\displaystyle{c_s=\sqrt{\frac{\gamma k_B T}{\mu m_H}}}$ and the
Alfv\'enic Mach number $M_a={V_0/c_a}$, where $c_a={B_x/(4\pi
\rho_0)^{1/2}}$ is the Alfv\'en speed of the unperturbed system.
The plasma beta is here defined by $\displaystyle{\beta={M_a^2\over
M_s^2}}$. 

These equations have exact solutions in the form of circularly
polarized plane Alfv\'en waves of arbitrary amplitude (Ferraro,
1955). They read $b_0=B_\perp\exp\left[ {-i\sigma (k_0 x-\omega_0 t)}\right]$
with constant density $\rho_0$ and temperature $T_0$,  and zero
longitudinal  velocity $u_0=0$. The transverse wave velocity 
is related to the magnetic field perturbation by
$\displaystyle{v_0=V_\perp \exp\left[ {-i\sigma (k_0 x-\omega_0
      t)}\right]=-\frac{b_0}{\sqrt{4\pi\rho_0}}}$. The polarization of 
the wave 
is determined by the parameter $\sigma$, with $\sigma=+1$ ($\sigma=-1$) for a
right-handed (left-handed) wave. In absence of dispersive effect we
can, without restriction, take $\sigma=-1$. Moreover, $B_\perp$ can
be taken real. The dispersion relation reads
$\omega_0^2 =c_a^2 k_0^2$.

\subsection{Ambipolar diffusion and wave steepening}
Before considering the idealized configuration of a completely
ionized medium with a circularly polarized monochromatic Alfv\'en
wave train of infinite length, it is important to estimate the
timescale of two important processes which are not considered in this 
study.
The first one is the ambipolar diffusion which takes place in weakly ionized 
gas and which induces energy dissipation due to the friction between  the 
neutrals and the ions. The second one is the steepening of the MHD
waves that takes place when the wave polarization is not perfectly
circular or when the wave packet is modulated.

\subsubsection{Timescale of the ambipolar diffusion}
The ambipolar diffusion time is given by (Shu 1992):
\begin{equation}
  t _{da} = {  4 \pi \gamma _{\rm da} \rho _n \rho _i L^2 \over  B^2  },
\end{equation}
where $\gamma _{\rm da}$ is the friction coefficient between ions of
density $\rho_i$  and neutrals of density $\rho_n$, 
$L$ is the typical scale to be considered and $B$ is the magnetic
intensity.

In the case of a molecular cloud, it has been estimated to 
$3 \times 10 ^{13} \, {\rm cm^{3} \, g ^{-1} \, s^{-1}}$ (Shu 1992). 
In the case of the
atomic gas, since the mass of the neutral is twice smaller, $\gamma _{\rm da}$
is twice larger, leading to
\begin{equation}
t _{da} \simeq 5 10^{15} \left( {  5 \mu G \over B } \right) ^2 \,
 { \xi _i \over 10^{-3}  } \,
 \left( { n _n \over  100 \, {\rm cm ^{-3} }   } \right)
 ^2 \, \left(L \over { 10^{18} \, {\rm cm}  } \right) ^2 {\rm s},
\end{equation}
where $\xi_i$ is the gas ionization and $n_n$ the particle density of the 
neutrals. 

In the CNM the ionization is about $4 \times 10 ^{-4}$, therefore at the scale
of say 0.1 pc, the ambipolar diffusion time is about $2\times 10^{14}$ s or about
$10 Myr$. In the WNM, the density is about 1 $cm^{-3}$, the ionization about 
0.1, therefore at a scale of say 1 pc, the ambipolar diffusion time is  
$5\times 10^{14}$ s. In both cases, ambipolar diffusion operates on
a rather long characteristic time and can thus safely be neglected.

\subsubsection{Timescale of the wave steepening}
The equation governing the 
dynamics of parallel propagating Alfv\'en waves in the long
wavelength, small amplitude limit was derived  by Cohen \& Kuslrud
(1974). A similar equation was derived previously by Rogister (1971)
from the Vlasov-Maxwell system in the context of a collisionless plasma, taking
into account kinetic effects such as dispersion and Landau
damping. The main point is that the evolution of the magnetic field
components perpendicular to the ambient field involves a
nonlinearity of the form $\alpha \partial_x \left ( |b|^2 b \right
)$, where $\alpha$ is a coefficient depending on $\beta$. From this
formula, one sees in particular that circularly polarized Alfv\'en waves, for
which $|b|$ is constant, do not steepen. In the case where the wave
amplitude is modulated, the characteristic time of steepening is
given by (using Eq. (40) in Cohen and Kulsrud 1974)
\begin{equation}
\tau=\frac{4}{3}\frac{1-\beta}{c_A}\frac{B_0^2}{\partial_x |b|^2}.
\end{equation}
Denoting by $\lambda$ the wavelength, by $l_c$ the typical length
of the amplitude modulation and by $\omega_s$ the characteristic
growth rate associated with wave steepening, one gets
\begin{equation}
\frac{\omega_s}{\omega_0}=
\frac{3}{4(1-\beta)}\frac{\lambda}{l_c}\frac{b_0^2}{B_0^2}.\label{steepening}
\end{equation}
It is to be noted that Cohen-Kulsrud equation is not valid for
$\beta$ close to unity, a point where sound waves have the same
phase speed as Alfv\'en waves and have thus to be retained
explicitely in the nonlinear dynamics.
In general however, we see that the ratio $\omega_s/\omega_0$
depends quadratically on the wave amplitude and is inversely
proportional to the normalized scale of amplitude modulation.
As an example, for a wave train of amplitude $b_0/B_0=0.5$ and
coherence length $l_c/\lambda=5$, with $\beta=0.5$, one gets 
$\omega_s/\omega_0=0.075$, a value about  five times smaller than
the typical growth rates  shown in Fig. 1.
This process of wave steepening thus appears subdominant with
respect to the combined thermal and decay instabilities considered
in this paper. Moreover, as will be demonstrated by the numerical
simulations, wave steepening does indeed occur but it does not
affect our conclusions.

Another process is at play in real three-dimensional situations,
namely turbulent cascade, which occurs mainly in
transverse directions as a result of nonlinear interaction between
counter-propagating wave packets. The nonlinear eddy turnover time is
expected to be smaller than the other processes mentioned above for
high enough amplitude and situations where Alfv\'en waves propagate
in both directions in roughly equal amount. When waves propagate in
a privileged direction, as in the solar wind, the turbulent cascade
is much less efficient, leaving enough time for the other
instabilities to develop.  The interaction between waves and
turbulence in compressible MHD is still a very debated topic (see
e.g. Cho \& Lazarian 2003). Three-dimensional simulations
will have to be performed to address the competition between all
these phenomena.

\subsection{Derivation of the dispersion relation}

We shall now consider perturbations about the exact solution
described above, namely a circularly polarized Alfv\'en wave 
propagating in a uniform gas at thermal equilibrium and
determine the dispersion relation for normal modes.
Such an analysis has been performed by Goldstein (1978) for an
isothermal gas and by Lou (1996) for a self-gravitating gas. 
Let us write the perturbations as

$\displaystyle{\rho = \rho _0 + \delta \rho , 
T= T _0 + \delta T , v = v_0  + \delta v , 
b= b_0 + \delta b , u = \delta u }$.

The linearized equations write, denoting by a star the complex conjugate,

\begin{eqnarray}
\nonumber
&&\partial _t \delta \rho + \rho _0 \partial _x \delta u = 0, \\
\nonumber
&&\rho _0 \partial _t \delta u = - \partial _x \delta P - {1 \over 8
\pi} \partial _x (b_0 \delta b^*    + b_0^* \delta b ), \\
&&\rho _0 \partial _t \delta v + \delta \rho \partial _t v_0 +
\rho_0 \delta u \partial_x v_0 = {1 \over 4 \pi} B_x
\partial _x \delta b, \\
\nonumber
&&\partial _t \delta b + \delta u \partial _x b_0 + b_0 \partial _x \delta u- B_x
\partial _x \delta v  = 0, \\
\nonumber
&&C _v( \partial _t \delta T + (\gamma -1) T_0 \partial _x \delta u) =
- ( \Big (\frac{\partial {\cal L}}{\partial \rho}\Big )_T \delta \rho+ \Big
(\frac{\partial {\cal L}}{\partial T}\Big )_{\rho}  \delta T ).  
\end{eqnarray}

Taking perturbations in the form
\begin{eqnarray}
\nonumber
\delta \rho &=& \widehat{ \delta \rho}  \exp(-i \omega t + i k x ) + c.c. , \\
\nonumber
\delta T &=&  \widehat{ \delta T} \exp(-i \omega t + i k x ) + c.c., \\
\delta u &=&  \widehat{ \delta u} \exp(-i \omega t +i  k x ) + c.c., \\
\delta v &=&  \widehat{ \delta v}^+ \exp(-i (\omega+\omega_0) t + i (k+k_0) x
) +   (\widehat{ \delta v}^{-})^* \exp(i (\omega-\omega_0) t - i (k-k_0) x
)  , \nonumber \\
\nonumber
\delta b  &=&  \widehat{ \delta b}^+ \exp(-i (\omega+\omega_0) t + i (k+k_0) x +
(\widehat{ \delta b}^-)^* \exp(i (\omega-\omega_0) t - i (k-k_0) x
),
\end{eqnarray}
we obtain the following equations
\begin{eqnarray}
\nonumber
-\omega \widehat{\delta  \rho} + k\rho _0  \widehat{\delta u} &=& 0, \\
\nonumber
-\rho _0 (\omega + \omega _0) \widehat{\delta v}^+ + (-\omega _0 \widehat{\delta
\rho} + k _0\rho _0 \widehat{\delta u} ) V _\perp &=& {B_x \over 4 \pi}
 (k+k _0) \widehat{\delta b}^+ , \\
\nonumber
-\rho _0 (\omega - \omega _0) \widehat{\delta v}^- + (\omega _0  \widehat{\delta
\rho} - k_0\rho _0 \widehat{\delta u} ) V _\perp &=& {B_x \over 4 \pi}
 (k - k _0) \widehat{\delta b}^-, \\
-\omega\rho _0  \widehat{\delta u}  + k {k _B \over\mu m_H} (\rho_0 \widehat{\delta
T} + T_0 \widehat{\delta \rho}) + {B _\perp \over 8 \pi } k
(\widehat{\delta b}^+ + \widehat{\delta b}^-)  &=& 0 , \\ 
\nonumber
-(\omega  + \omega_0) \widehat{\delta b}^+ + B _\perp(k+k _0) \widehat{\delta u}  
- B _x (k +k_0) \widehat{\delta v}^+  &=& 0, \\
\nonumber
-(\omega  - \omega_0) \widehat{\delta b}^- +B_\perp(k-  k _0)\widehat{\delta u}
- B_x (k - k_0) \widehat{\delta v}^- &=& 0, \\
\nonumber
i C_v ( -\omega \widehat{\delta T} + (\gamma -1) T _0 k \widehat{\delta u}) &=& 
- \left( \partial _\rho {\cal L} \widehat{\delta \rho} + \partial _T {\cal L}
\widehat{\delta T} \right) .
\end{eqnarray}

The fluctuations $\widehat{\delta T}$, $\widehat{\delta \rho}$, and
 $\widehat{\delta b}^{\pm}$ can easily be 
 obtained as a function of $\widehat{\delta u}$ in the form
\begin{eqnarray}
\nonumber
\omega \widehat{\delta \rho} &=& k \rho _0  \widehat{\delta u}, \\
\widehat{\delta T} &=& \frac{\widehat{\delta u}}{- \partial _T {\cal
    L} + i C_v \omega} \left( ik C_v (\gamma -1) T _0  + 
 {k \over \omega }\rho _0 \partial _\rho {\cal L} \right), \\
\nonumber
\widehat{\delta b}^\pm  &=& { \widehat{\delta u} \over  \omega _\pm ^2 - c_a^2 k
_\pm ^2} B _\perp k _\pm { \omega _0 ^2 k \over \omega k _0}
\left(  { \omega _\pm \over \omega _0} {\omega  \over \omega _0 } {k _0
\over k}  \pm {\omega \over \omega _0} {k _0 \over k } - \pm 1   \right). 
\end{eqnarray}

One then gets the dispersion relation  
\begin{eqnarray}
\nonumber
\left( -\omega ^2 + k^2 {k_B \over \mu m_H} T _0 +  k^2 {k_B \over\mu m_H} 
\left( { i C_v (\gamma -1) T_0 \omega + \rho _0\partial_\rho {\cal L} 
\over - \partial _T {\cal L} + i C _v \omega  } \right)  \right) & 
\left( {\omega\over\omega _0} - {k \over k_0}  \right) 
\left( \left({\omega\over\omega_0} + {k \over k_0}  \right)^2 - 4  \right) = \\ 
-\omega _0 ^2 \left( { B _\perp \over B_x} \right)^2 
\left( {k \over  k_0} \right) ^2 
\left( \left({\omega \over \omega _0}\right)^3
+ {k \over k _0} \left( {\omega \over \omega _0} \right)^2
 -3 {\omega \over \omega _0} + {k \over k _0} \right) , &
\label{dispersion}
\end{eqnarray}
which can be rewritten

\begin{eqnarray}
 \widetilde{\omega}^2-\beta \widetilde{k}^2
\frac {\widetilde {\omega} + i \widetilde {\omega_b}} {\widetilde{\omega}+
    i \widetilde {\omega_c} }  =  
 \widetilde{k}^2 A^2 \frac {\widetilde{\omega}^3 + \widetilde{k}
   \widetilde{\omega}^2 -3 \widetilde{\omega} + \widetilde{k} } 
{(\widetilde{\omega} - \widetilde{k} ) ( (\widetilde{\omega} + \widetilde{k} )^2 - 4)},
\label{disp_rel}
\end{eqnarray}
where
$\displaystyle{\widetilde{\omega_c}=\frac{\omega_c}{\omega_0}=\frac{\partial_T
    {\cal L}}{C_v\omega_0}}$,  
$\displaystyle{\widetilde{\omega_b}=\frac{\omega_b}{\omega_0}=\frac{
    T_0\partial_T {\cal 
      L}-\rho_0\partial_{\rho} 
 {\cal L} }{\gamma C_v T_0 \omega_0}}=\frac{1}{\gamma C_v\omega_0}
\Big (\frac{\partial {\cal L}}{\partial T}\Big )_P$,
$\widetilde{\omega}=\omega/\omega_0$, $\widetilde{k}=k/k_0$ and
$A=B_\perp/B_x$. 

Note that in the limit  $A=0$, Eq.~(\ref{disp_rel}) becomes identical
with the dispersion relation obtained by  
Field (1965) whereas if ${\cal L}=0$, it reduces to the
dispersion relation obtained by Goldstein (1968).  

\subsection{Asymptotic behaviors}

Before numerically solving  Eq.~(\ref{disp_rel}), we consider various
asymptotic limits.

\subsubsection{Static magnetic field}
It is possible to recover the dispersion relation in the absence of
waves for a situation where the ambient field is oblique, making an
angle $\theta$ with the 
$x$-axis. In this case $B_x=B_0\cos\theta$ and one must take the
limit $\omega_0\rightarrow 0$ with
$\displaystyle{\frac{\omega_0}{k_0}=c_A=\frac{B_0}{\sqrt{4\pi\rho_0}}\cos\theta}$ and
$A=\tan \theta$. It follows that
\begin{equation}
\omega^2- k^2 c_s^2
\frac{\omega+i\omega_b}{\omega+i\omega_c}=
\frac{ k^2\omega^2v_A^2\sin^2\theta}{\omega^2-k^2v_A^2\cos^2\theta},
\label{bstat}
\end{equation}
where we denote $\displaystyle{v_A^2=\frac{B_0^2}{4\pi\rho_0}}$.

In the case  $\theta=\pi/2$ and close to the  threshold ($\omega \simeq 0 \ll
kc_s$), we have
\begin{equation}
 \omega \simeq -i \frac{v_A^2\omega_c+c_s^2\omega_b}{v_A^2+c_s^2}.
\label{cas3}
\end{equation}
Since the magnetic field is purely transverse,  the
magnetic tension vanishes and the magnetic pressure adds up to the
thermal pressure making the gas more stable. The criterion for thermal instability
is simply $v_A^2\omega_c+c_s^2\omega_b \le 0$ which was first obtained by Field (1965).
It shows that a purely transverse magnetic field can suppress the thermal instability.
This is because, in this geometry, the magnetic pressure is
proportional to the square of the density and
therefore $\partial _\rho P _{\rm tot}  $  can be positive even if
$\partial _\rho P _{\rm therm}$  is negative.

\subsubsection{Instability thresholds}
This section addresses the neighborhood of the instability,
a situation where $\Im(\omega) \rightarrow 0$.

We shall first discuss the case $\Re(\omega) = 0$, corresponding to the
so-called condensation or entropy mode. 
In the situation where $|\widetilde{\omega_b}|$ and
$|\widetilde{\omega_c}|$ are smaller than $\widetilde{k}$, the characteristic
cooling time is longer than the Alfv\'en crossing time and
the dispersion relation reduces to  
\begin{equation}
\widetilde{\omega}  \simeq- i\frac{\beta
  (4-\widetilde{k}^2)\widetilde{\omega_b}+A^2\widetilde{\omega_c}}
{\beta(4-\widetilde{k}^2)+A^2}.\label{disp}
\end{equation} 
When $\beta$ is not too small sound waves have time to restore
pressure equilibrium while the gas cools and one expects that the
growth rate will be close to the isobaric one. In the opposite case
where $\beta$ is very small, the growth rate should be close to the
isochoric one since Alfv\'en waves are not accompanied by pressure
or density perturbations.
These conclusions are easily recovered form Eq. (\ref{disp}) which
for very small amplitude ($A^2/\beta(4-\widetilde{k}^2) \ll 1$) gives 
\begin{equation}
\widetilde{\omega} \simeq -i \Big(\widetilde{\omega_b}+
\frac{A^2}{\beta(4-\widetilde{k}^2)}\Big (\widetilde{\omega_c}
-\widetilde{\omega_b} \Big) \Big ),
\end{equation} 
while, in the case where $\beta(4-\widetilde{k}^2)/A^2 \ll 1$ it rewrites
\begin{equation}
\widetilde{\omega} \simeq -i \Big(\widetilde{\omega_c}+
\frac{\beta(4-\widetilde{k}^2)}{A^2}\Big (\widetilde{\omega_b}
-\widetilde{\omega_c} \Big) \Big ).
\end{equation} 

In the case where $A^2/\beta \ll 1$, more likely to be met in the ISM,  the
effect of the Alfv\'en wave 
depends both on the sign of $\widetilde{\omega_c}
-\widetilde{\omega_b}$ and of that of $\widetilde{k}- 2$.
In a typical region of the ISM with T$\simeq$ 1000 K,
$\widetilde{\omega_c}-\widetilde{\omega_b} \ge 0 $ 
and therefore the waves stabilize (destabilize) the gas if $\widetilde{k} < 2$,
(respectively $\widetilde{k} \ge 2$). 

In the limit  $\widetilde{k} \gg 1 $ and/or for $A=0$,  
one recovers the isobaric growth rate, the destabilizing effect of
the Alfv\'en waves  becoming asymptotically small as $\widetilde{k}$ increases.

If $\widetilde{\omega_b}\ll\widetilde{k} \ll 1 $, we find 
a growth rate similar to that given by Eq.~(\ref{cas3}) except
for the factor 2 that divides  $v_A^2$. This is due to the magnetic tension 
that tends to unbend the magnetic field lines making the
stabilization of the magnetic pressure less efficient.

Two different limits are obtained when $\widetilde{k}
\ll\widetilde{\omega_b}$. When the isochoric criterion is not
verified and $\widetilde{\omega}/\widetilde{k}\equiv\alpha$ remains
finite, a situation where the 
growth rate vanishes with $\widetilde{k}$, one gets
\begin{equation}
(1-\alpha)(\alpha^2-\beta\frac{\widetilde{\omega_b}}{\widetilde{\omega_c}})= 
(1-3\alpha)\frac{A^2}{4}. \label{sk}
\end{equation}
In the absence of waves, the growth rate asymptotically approaches
$\displaystyle{\widetilde{\omega}=
\pm\beta^{\frac{1}{2}}\widetilde{k}(\widetilde{\omega_b}/
	\widetilde{\omega_c})^ {\frac{1}{2}}}$  (Meerson 1996). 
Equation (\ref{sk}) shows in general, assuming $\alpha$ and
$A$ small and thus $\displaystyle{\alpha^2\approx \beta\widetilde{\omega_b}/
\widetilde{\omega_c}+A^2/4}$, that the Alfv\'en waves have a stabilizing effect.

When $\widetilde{\omega_c}<0$, the growth rate for $\widetilde{k}
\ll|\widetilde{\omega_c}|$ remains finite, equal to
$\widetilde{\omega}=-i\widetilde{\omega_c}$, independently of the
presence of Alfv\'en waves.

Thus, in the limit of very short and long wavelengths, Alfv\'en
waves do not modify the growth rates of the  
thermal instability obtained by Field (1965).

For completeness, we now consider the case of adiabatic
perturbations, corresponding to a wave mode with $\Re(\omega)
\ne 0 $. For this
purpose, we restrict ourselves to the case 
of low amplitude waves, i.e $A \ll 1$ and we set
$\widetilde{\omega} =\widetilde{\omega} _r  + i \widetilde{\omega}_i$. 
We therefore have $\widetilde{\omega}_r \gg
\widetilde{\omega}_i$. Moreover, it is assumed that 
$\widetilde{\omega_c}\ll 1$ and $\widetilde{\omega_b}\ll 1$ so that
they can be neglected when multiplied by $\widetilde{\omega}_i$.
With these assumptions, one finds  that to the first order
\begin{eqnarray}
\widetilde{\omega}  &=& \beta^{\frac{1}{2}}\widetilde{k} 
+\frac{i}{2}(\widetilde{\omega_b}-\widetilde{\omega_c})
+\frac{A^2}{2\beta}\Big (\frac{\beta^{\frac{1}{2}}\widetilde{k}
  +i\widetilde{\omega_c}}{\beta^{\frac{1}{2}}
  -1}\Big )\Big(\frac{(\beta^{\frac{3}{2}} +\beta)\widetilde{k}^2
  +1-3\beta^{\frac{1}{2}}} {(1+\beta^{\frac{1}{2}})^2\widetilde{k}^2-4}\Big ).
\label{sonic_crit}
\end{eqnarray}
The real part, $\omega _r \simeq C_s k$ simply  corresponds to  a sonic wave.

When $A=0$, one finds that the conditions for thermal stability of a
 sonic wave is
$\widetilde{w _c} - \widetilde{w _b} \ge 0$, a criterion already
 obtained by Field (1965). 

When $A \ne 0$, the stability criterion is modified according to 
Eq.~(\ref{sonic_crit}), showing that the effect of the waves 
depends  in a complex way on $\beta$ and $\widetilde{k}$.

\subsection{Growth rate and physical discussion}\label{sec:rel}

\setlength{\unitlength}{1cm}
\begin{figure}
\begin{picture}(0,12)
\put(0,9){\includegraphics[width=7cm]{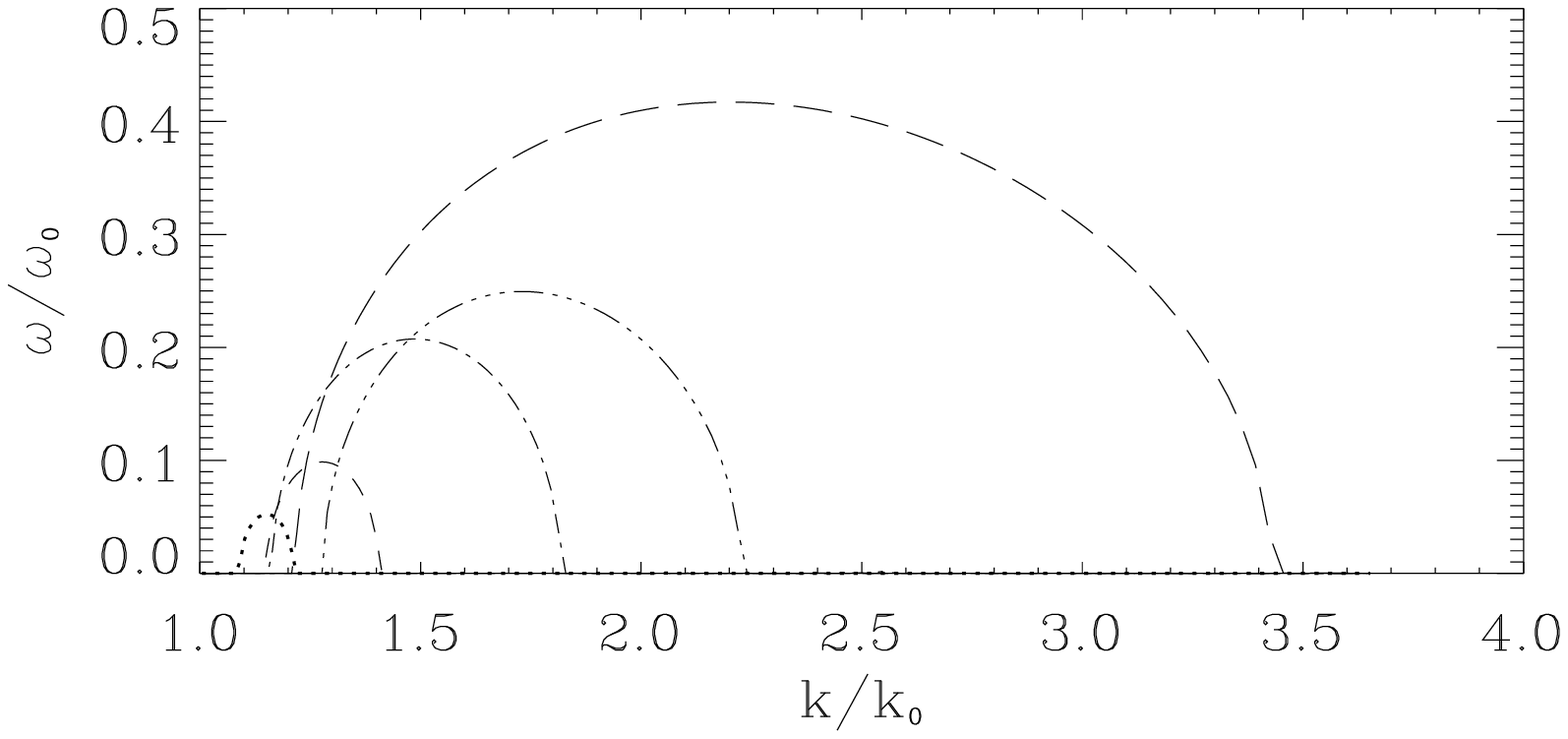}}
\put(0,6){\includegraphics[width=7cm]{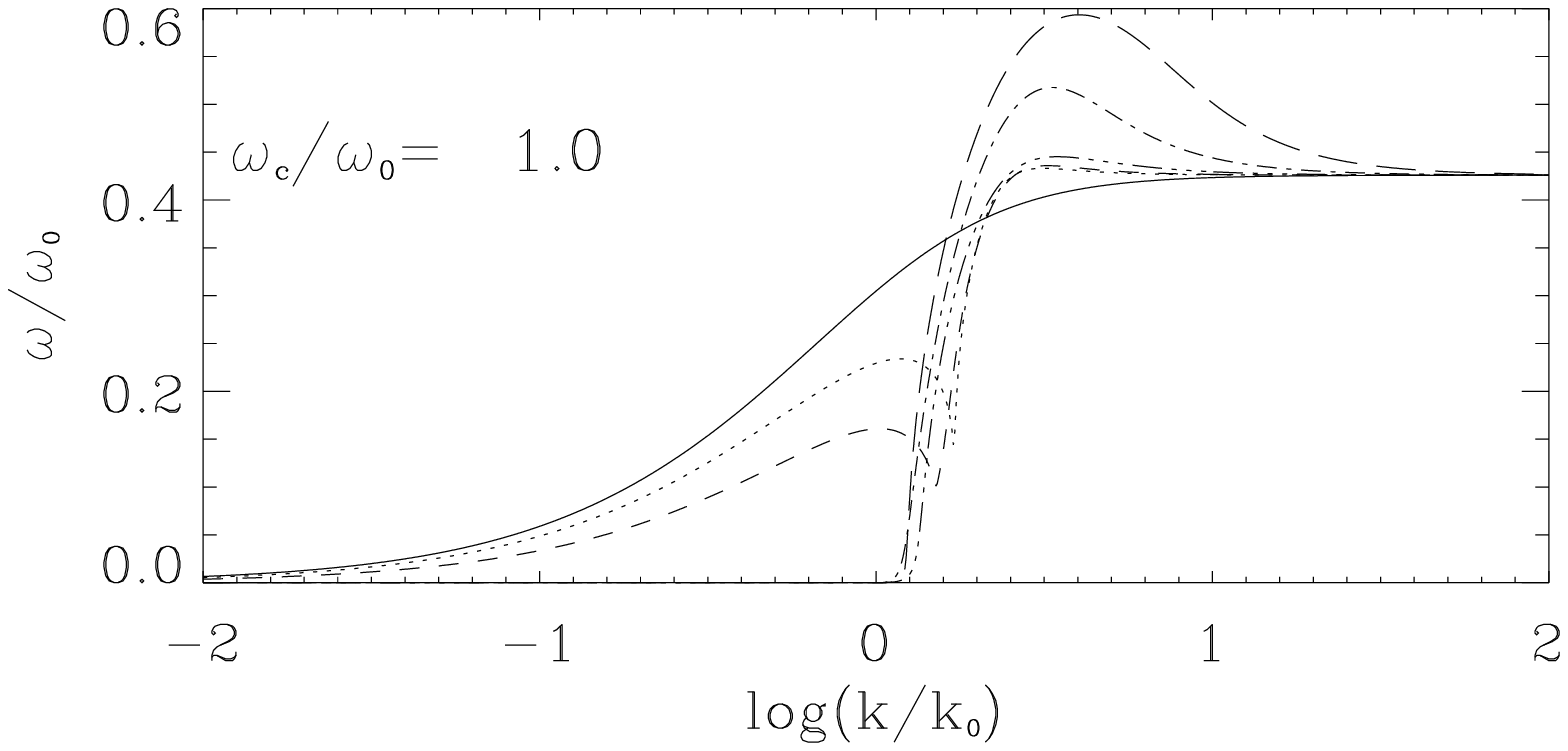}}
\put(0,3){\includegraphics[width=7cm]{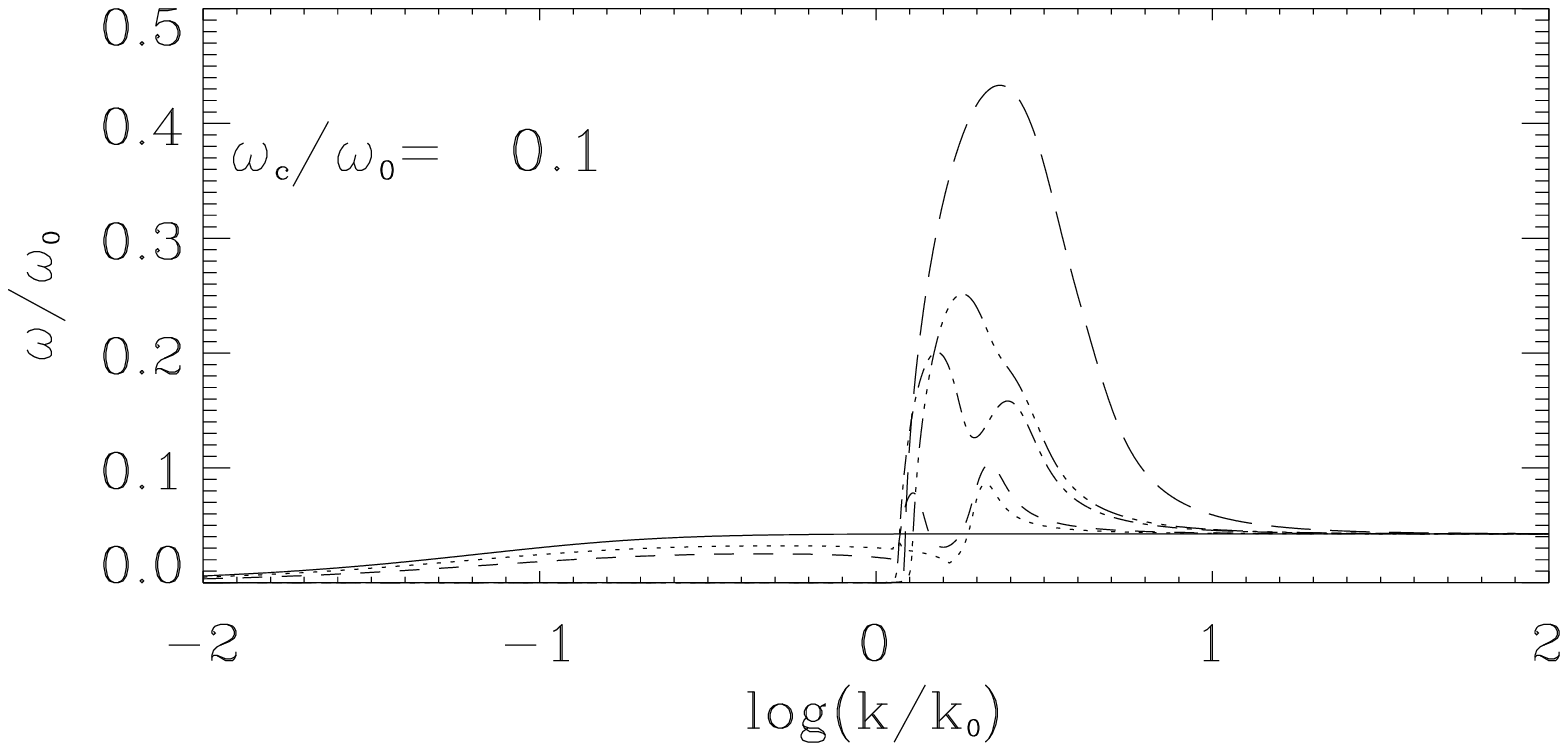}}
\put(0,0){\includegraphics[width=7cm]{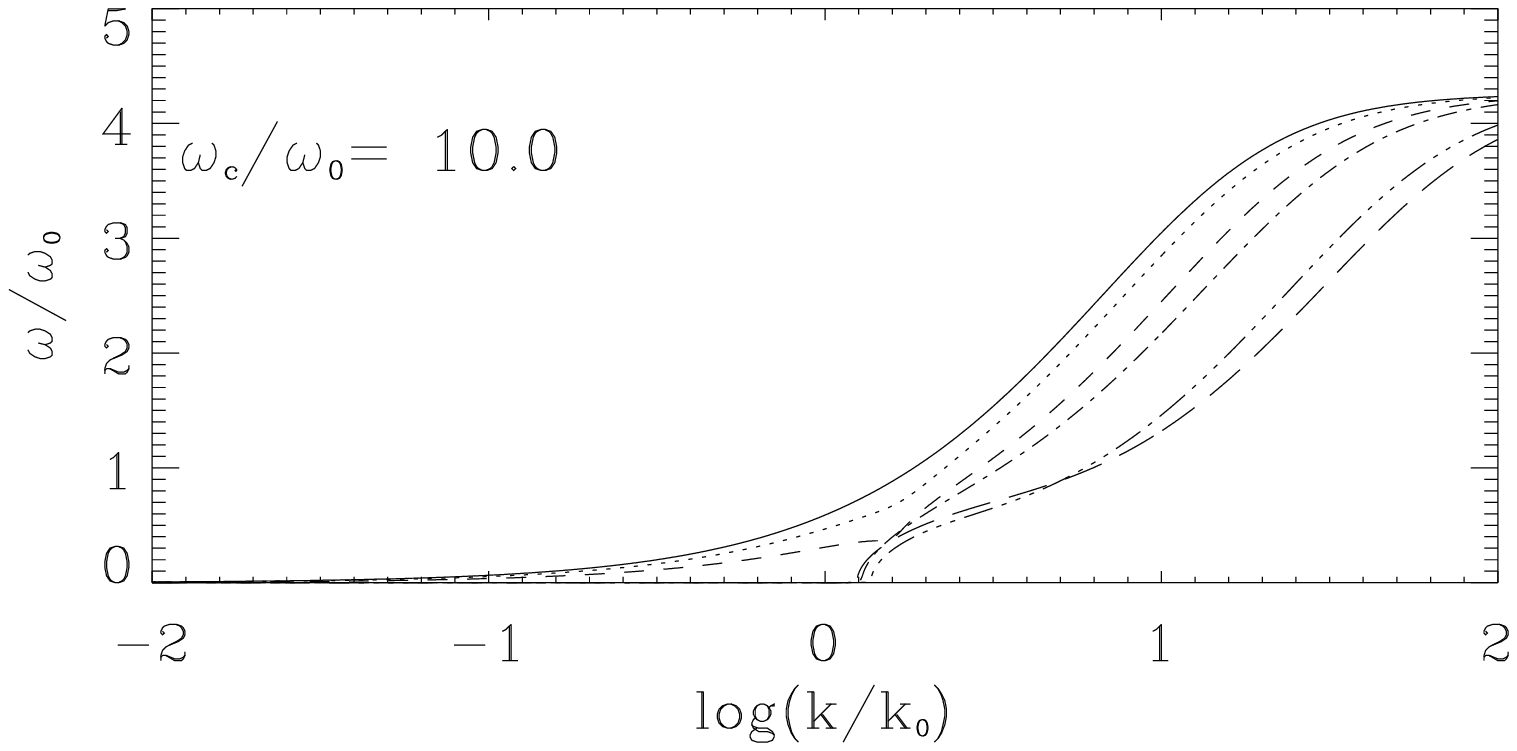}}
\end{picture}
\caption{Growthrate as a function of the wave number.
First panel is for the adiabatic case, second, third and fourth panels
display results for various values of $\omega _c / \omega _0$.
Solid line corresponds to the case $A=0$, dotted line to $\beta=0.9$, $A=0.5$, 
 short dashed line to $\beta=0.5$, $A=0.5$,  
dot-dashed line to $\beta=0.5$, $A=1$, 
double dot-dashed line to $\beta=0.1$, $A=0.5$ and 
long dashed line to  $\beta=0.1$, $A=1$. }
\label{growthrate}
\end{figure}

We now numerically solve Eq. (\ref{disp_rel}) 
which can be rewritten as a sixth order polynomial
containing 4 parameters to be specified,
namely $\beta$ and $A$  (characterizing the Alfv\'en wave)  and 
$\widetilde{\omega}_b$ and $\widetilde{\omega} _c$ (function of the
thermal processes).   
The value of $\widetilde{\omega} _c = \omega _c / \omega _0 $
represents the ratio of the temporal period  of the Alfv\'en wave divided by 
the cooling time  and  can be arbitrarily chosen. However, once
$\widetilde{\omega} _c$ is specified,   $\widetilde{\omega} _b$ depends on the 
thermal function. In order to estimate this parameter we use the standard cooling 
function of the neutral atomic ISM 
(Wolfire et al. 1995, 2003) which is used in Audit \& Hennebelle
(2005) in the thermally unstable 
 regime ($n=3 \, $cm$^{-3}$ and $T \simeq 500$ K).
To obtain the dispersion relation we integrate
Eq.~(\ref{dispersion}) for  $k/k_0$ between 0.01 and 100  using
logarithmic spacing.  The roots of the polynomials  are obtained
using the {\it zroots} subroutine (Press et al. 1992). 
 Here, we restrict our attention to the unstable branch only. 
In order to verify our method, we have reproduced the dispersion
relation for the decay instability of an isothermal gas presented in
Goldstein (1978). 

Fig.~\ref{growthrate} displays the results of the numerical
integration of Eq.~(\ref{dispersion}).  
First panel shows the adiabatic case for $\beta=0.9$, $A=0.5$
(dotted line), $\beta=0.5$, $A=0.5$ (short dashed line),  
$\beta=0.5$, $A=1$ (dot-dashed line), $\beta=0.1$, $A=0.5$ (double
dot-dashed line),  $\beta=0.1$, $A=1$  
(long dashed line).
As expected the circularly polarized Alfv\'en wave is unstable (decay
instability)  in a range of $k$  extending to a few times
$k_0$. Both the growth rate and the largest  
unstable value of $k$, increase with $\beta^{-1}$ and $A$.

The second panel shows results for $\widetilde{\omega} _c = \omega _c
/ \omega _0 = 1$. 
The full line corresponds to the hydrodynamical case 
whereas the others correspond to the same values of $\beta$ and $A$
as in the first panel. Various interesting features can be seen.
 i) When $k \rightarrow 0$ the effect of the Alfv\'en wave is to
decrease the growth rate and therefore to stabilize these modes with
respect to thermal instability. For $\beta =0.5, A=1$ and for
$\beta=0.1$, the modes whose wavenumber is smaller  than $k \simeq
1$ are perfectly stable. 
ii) The intermediate modes (i.e $k \simeq k_0$) are more unstable when
$\beta$ is smaller and $A$  is  higher. This is due to the decay
instability that 
the wave undergoes for these values of $k$. iii) When $k \rightarrow
\infty$,  the growth rate is independent of $\beta$ and $A$.
These features are in good agreement with the asymptotic limit
$\omega \simeq 0$ studied in the previous section.
Cases with a larger $\beta$, namely 1 and 1.5, have also been
explored but are not exposed here for conciseness since they are not  
directly relevant for the regions of the ISM we consider in this paper.
Although the decay instability 
 disappears for $\beta \ge 1$ (for small enough amplitude), we find no
 qualitative difference with the cases  $\beta \le 1$.
In particular the intermediate wavelengths are still
destabilized by the waves. This is in 
good agreement with the analytical study of the instability
threshold  (see Eq.~(\ref{disp})). 
 
The third panel shows results for  $\widetilde{\omega} _c = 0.1$. In
this case the cooling time is 10 times larger than the period of the
Alfv\'en waves. The dispersion relation is qualitatively similar to
the previous case. Quantitatively however, the intermediate wavelengths
are much more unstable than the small wavelengths. This is due to the
fact that in this range of parameters, the cooling time being larger
than the dynamical time of the waves, the fastest growing
instability is  the decay instability. Note that since
$\omega _0$ is 10 times larger in this case than in the case displayed
in panel 2, the value of $\omega / \omega _0$  in the hydrodynamical
case (full line)  is 10 times lower.

Fourth panel shows results for $\widetilde{\omega} _c = 10$. In this
case, the growth rate of the thermal instability is much shorter than the growth rate  of the decay
instability. Therefore  the only effect of the  Alfv\'en wave is to stabilize
the gas with respect to thermal instability.  This effect
increases when $\beta$ decreases and when $A$ increases.

These  results suggest that  the presence of non-linear circularly
polarized Alfv\'en waves in a thermally unstable medium like the
neutral interstellar atomic gas can have two
main effects. If the cooling time is short with respect to the temporal period
of the waves, then the waves stabilize the gas. Therefore  the gas 
can survive longer in the thermally unstable domain leading possibly
in the ISM, to a larger fraction of thermally unstable gas.
Since the short  wavelengths are the most unstable 
the trend is that the CNM is very fragmented into
several clouds. If the cooling time is larger than the wave
period, then the decay instability makes the intermediate modes 
($k \simeq k _0$) more unstable. 
In that case, the fraction of thermally unstable gas is not necessarily
larger (depending on $\beta$ and $A$) but
the CNM should be fragmented in structures having a
size of about $1/(k _0 \times \xi)$, where $\xi$ is the density ratio
between the CNM and the WNM.

\subsection{Effect of thermal diffusivity}
\setlength{\unitlength}{1cm}
\begin{figure}
\begin{picture}(0,9)
\put(0,6){\includegraphics[width=7cm]{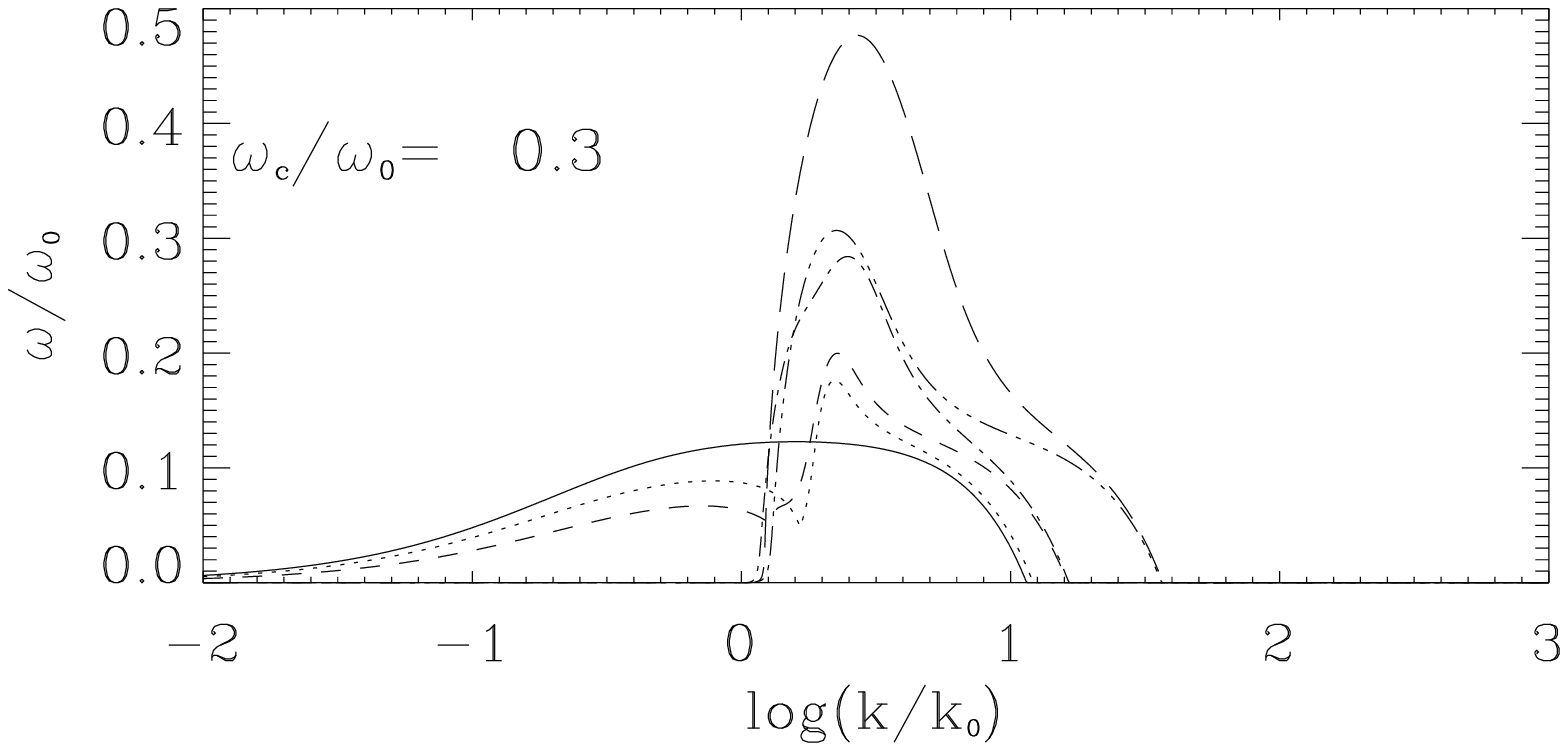}}
\put(0,3){\includegraphics[width=7cm]{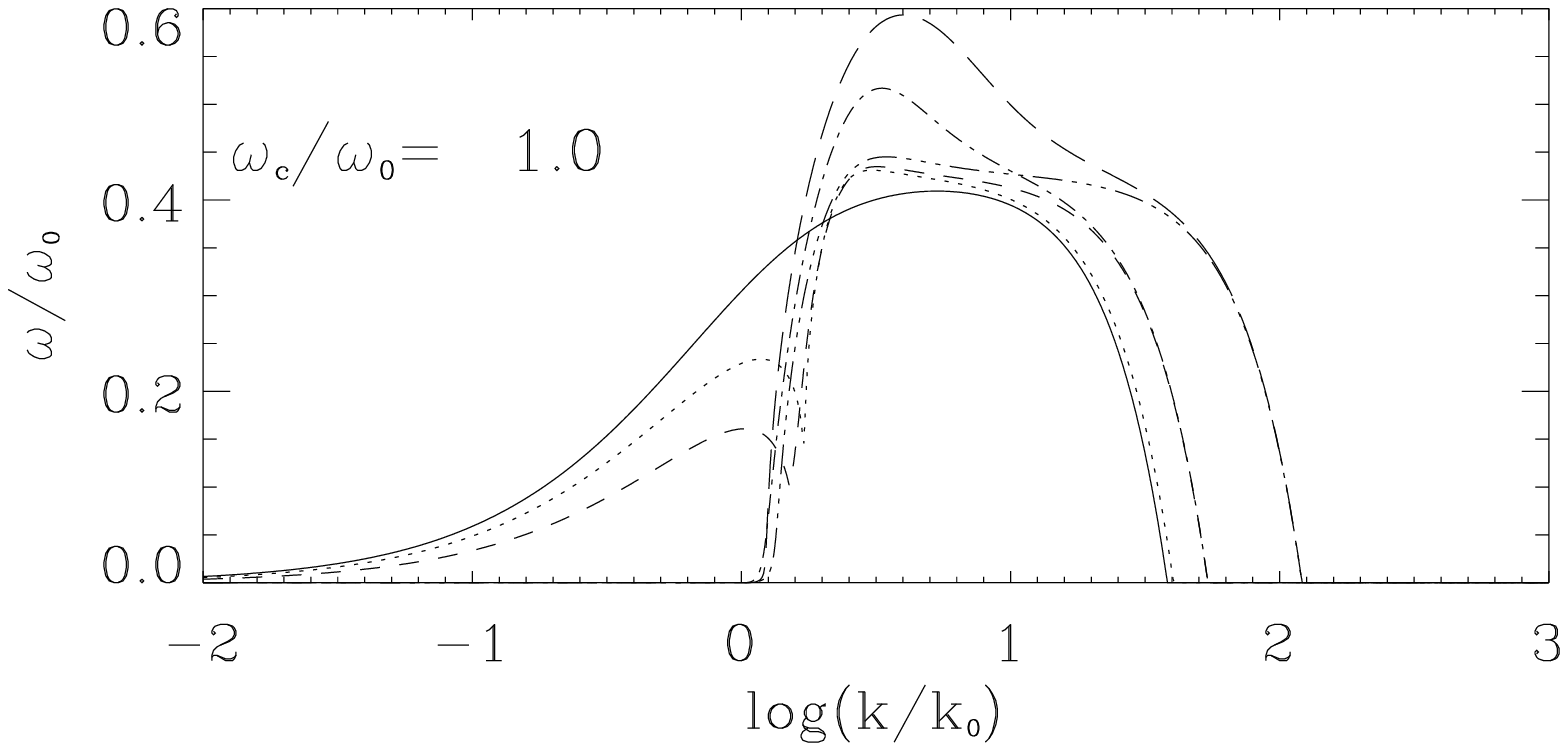}}
\put(0,0){\includegraphics[width=7cm]{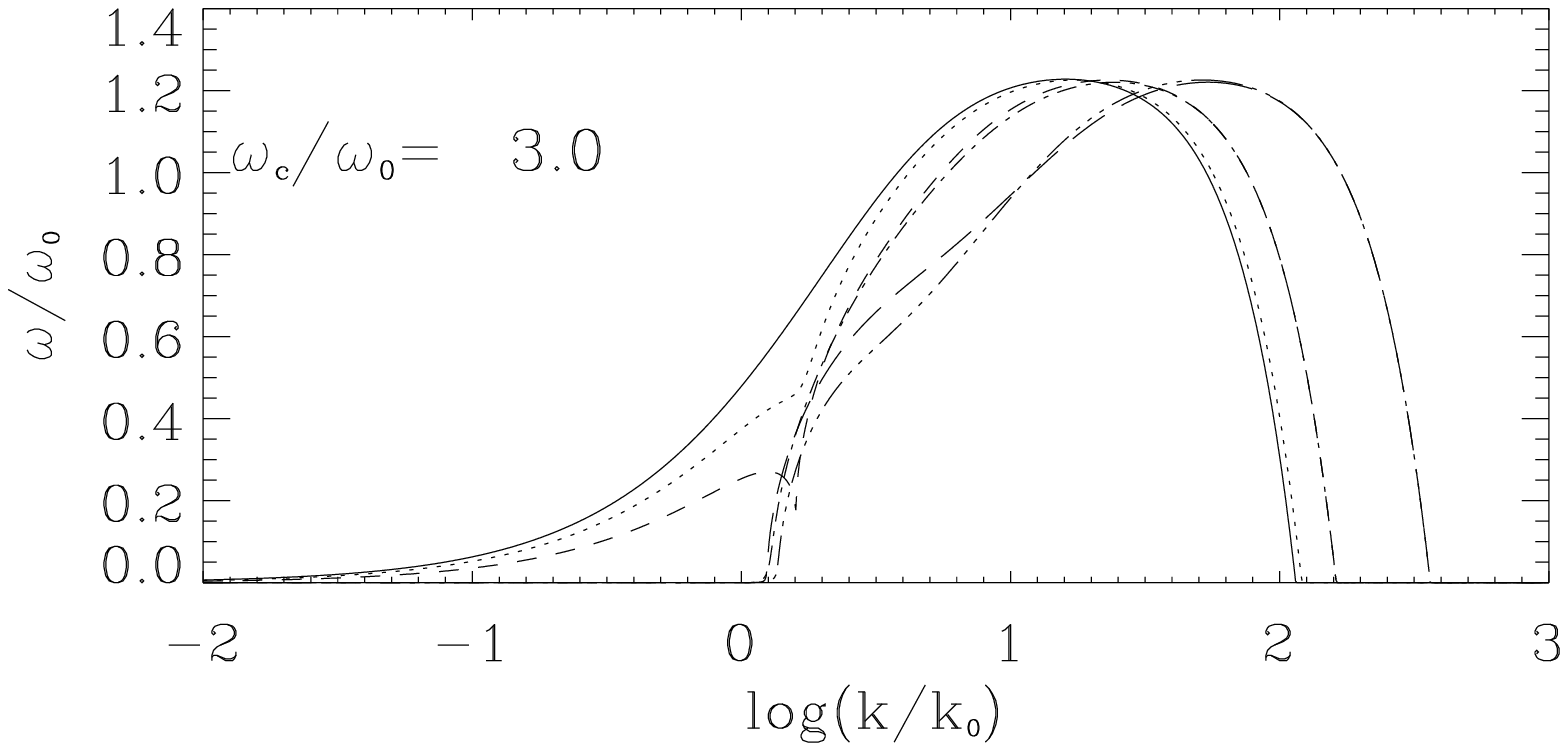}}
\end{picture}
\caption{Growthrate as a function of the wave number when the effect of the thermal diffusivity is 
taken into account. The curve styles are associated with the same
values of $\beta$ and $A$ as in Fig.~\ref{growthrate}.}
\label{growthrate_cond_therm}
\end{figure}

Here  we briefly consider the effect of thermal diffusivity.
When this term is taken into account, the dispersion relation (\ref{disp_rel})
becomes:
\begin{eqnarray}
 \widetilde{\omega}^2-\beta \widetilde{k}^2
\frac {\widetilde {\omega} + i (\widetilde {\omega_b} + \widetilde{\kappa_b} \widetilde{k}^2 ) } {\widetilde{\omega}+
    i (\widetilde {\omega_c} + \widetilde{\kappa_c} \widetilde{k}^2)}  =  
 \widetilde{k}^2 A^2 \frac {\widetilde{\omega}^3 + \widetilde{k}
   \widetilde{\omega}^2 -3 \widetilde{\omega} + \widetilde{k} } 
{(\widetilde{\omega} - \widetilde{k} ) ( (\widetilde{\omega} + \widetilde{k} )^2 - 4)},
\label{disp_rel_therm}
\end{eqnarray}
where $\widetilde{\kappa_c} = \kappa (T_0) \omega _0 \beta / (C_v \rho_0 C_s^2)$ and 
$\widetilde{\kappa_b} = \widetilde{\kappa_c} / \gamma$.

Fig.~\ref{growthrate_cond_therm} displays the growth rate obtained
from Eq. (\ref{disp_rel_therm}) using the fiducial value of $\kappa
(T_0)$ given at the beginning of Sect. 2 and for $\omega_c=0.3, \; 1, \; 3$. 
The different curves are associated with the same values of $\beta$
and $A$ as in Fig. (\ref{growthrate})
For small and intermediate values of $\widetilde{k}$ the shape and the values of $\widetilde{w}$ are very 
similar to the case of vanishing thermal conductivity.  
As expected however, thermal conduction introduces a cut-off at
small scale (Field 1965).  
The value of $\widetilde{k}$ for which $\widetilde{\omega}$ vanishes depends on $\beta$ and increases 
(by a factor 2 to 3 for the values considered here) when $\beta$
decreases, confirming the trends inferred previously, i.e.  the CNM should be more fragmented in the presence of
Alfv\'en waves leading to smaller CNM structures.

\section{Numerical study}\label{sec:simu}

\begin{figure}
\begin{picture}(0,13)
\put(0,9){\includegraphics[width=7cm]{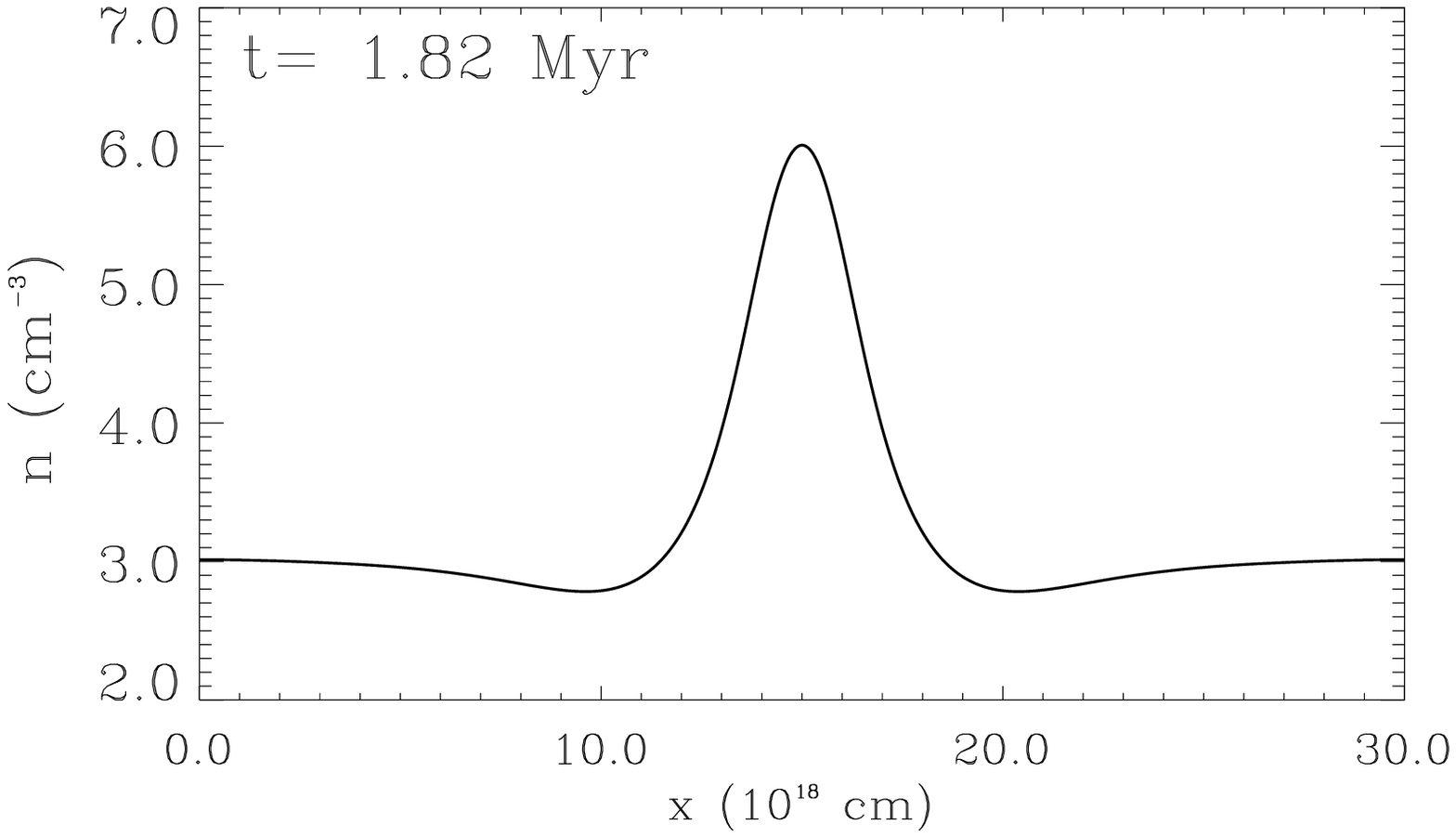}}
\put(0,5){\includegraphics[width=7cm]{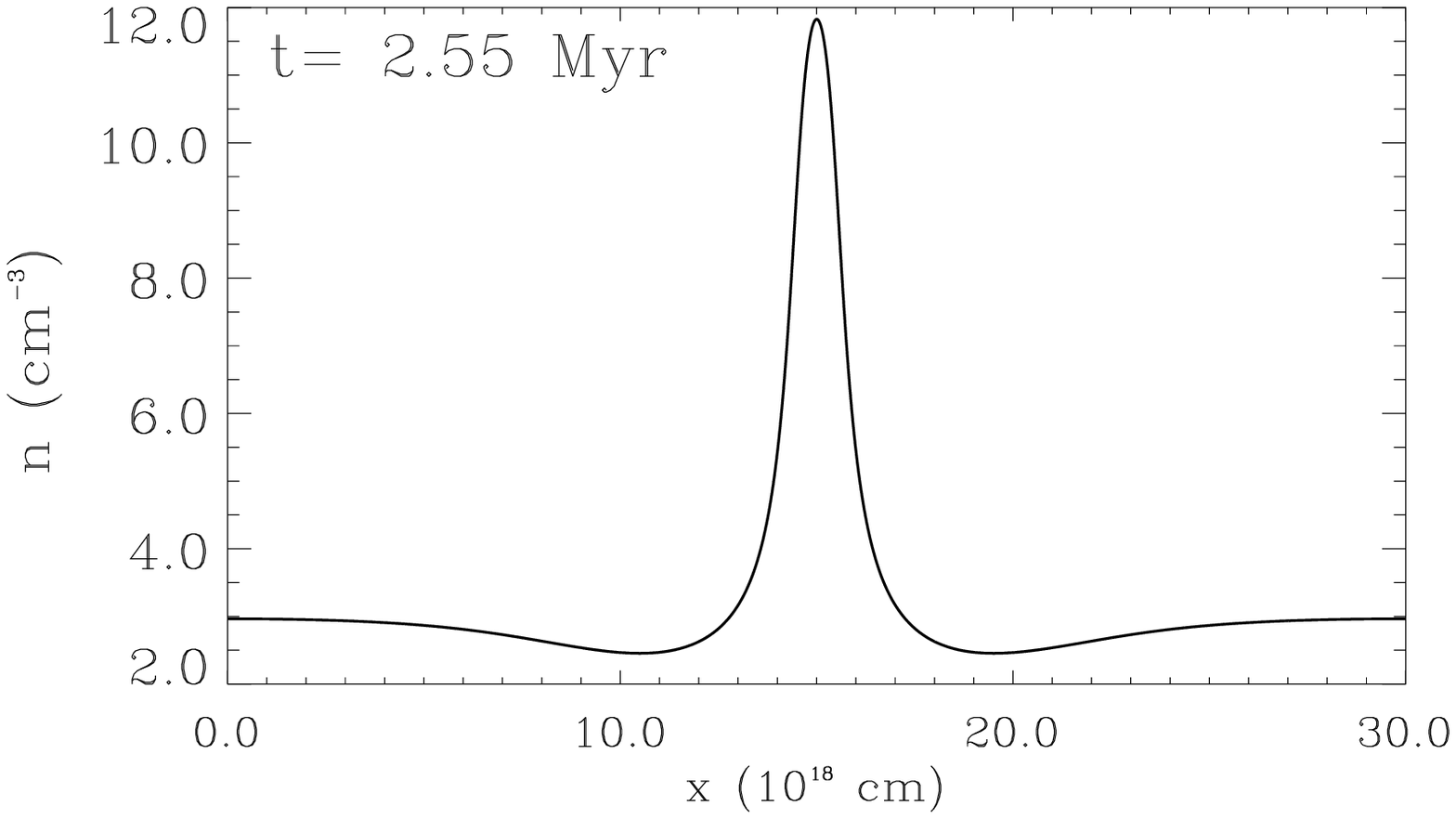}}
\put(0,1){\includegraphics[width=7cm]{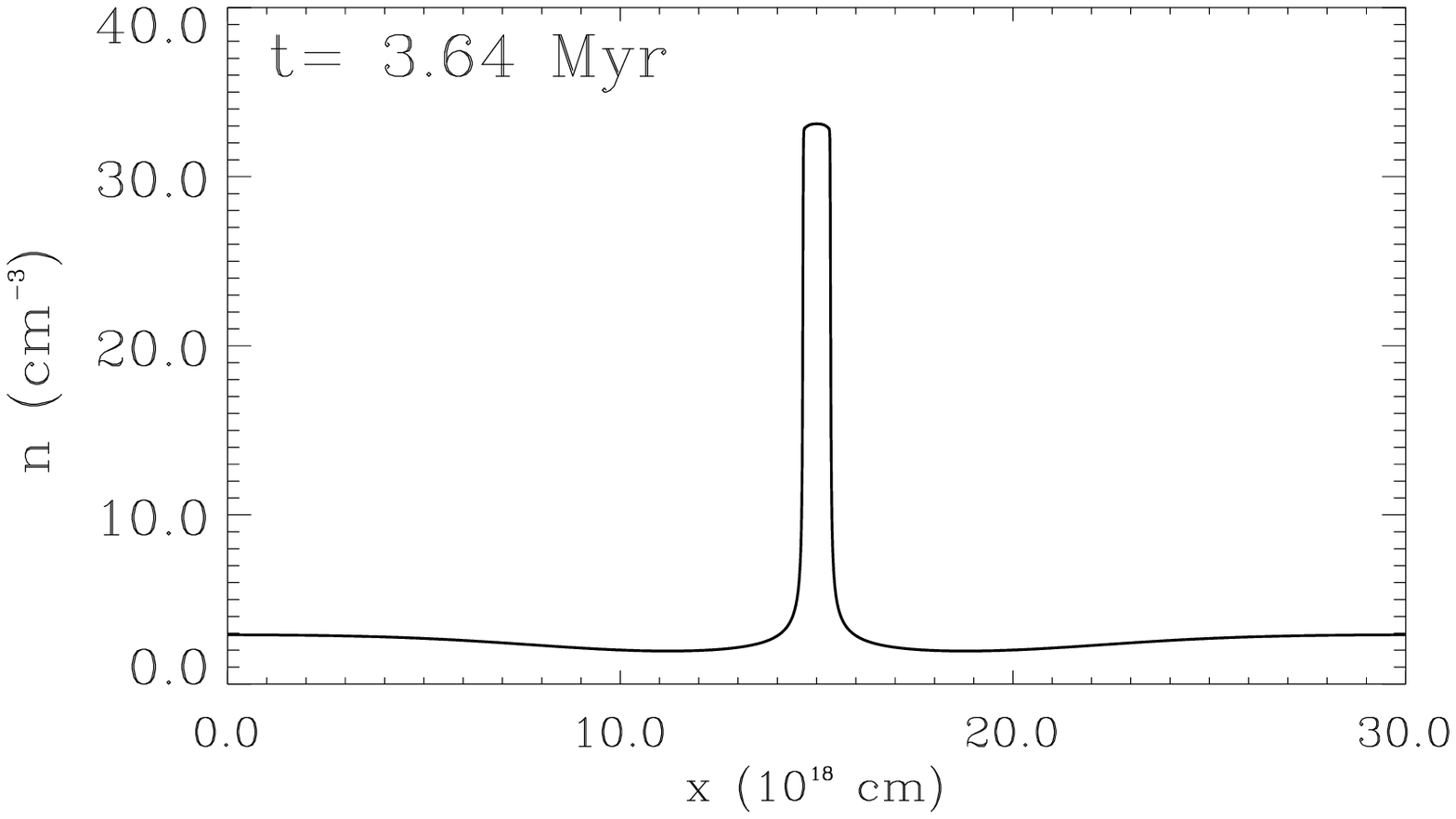}}
\end{picture}
\caption{Density field for 3 snapshots illustrating the development 
of the thermal instability in the hydrodynamical case. The density of the
perturbation increases until the gas reaches thermal equilibrium. }
\label{hydro}
\end{figure}

\begin{figure}
\begin{picture}(0,25)
\put(0,17){\includegraphics[width=7cm]{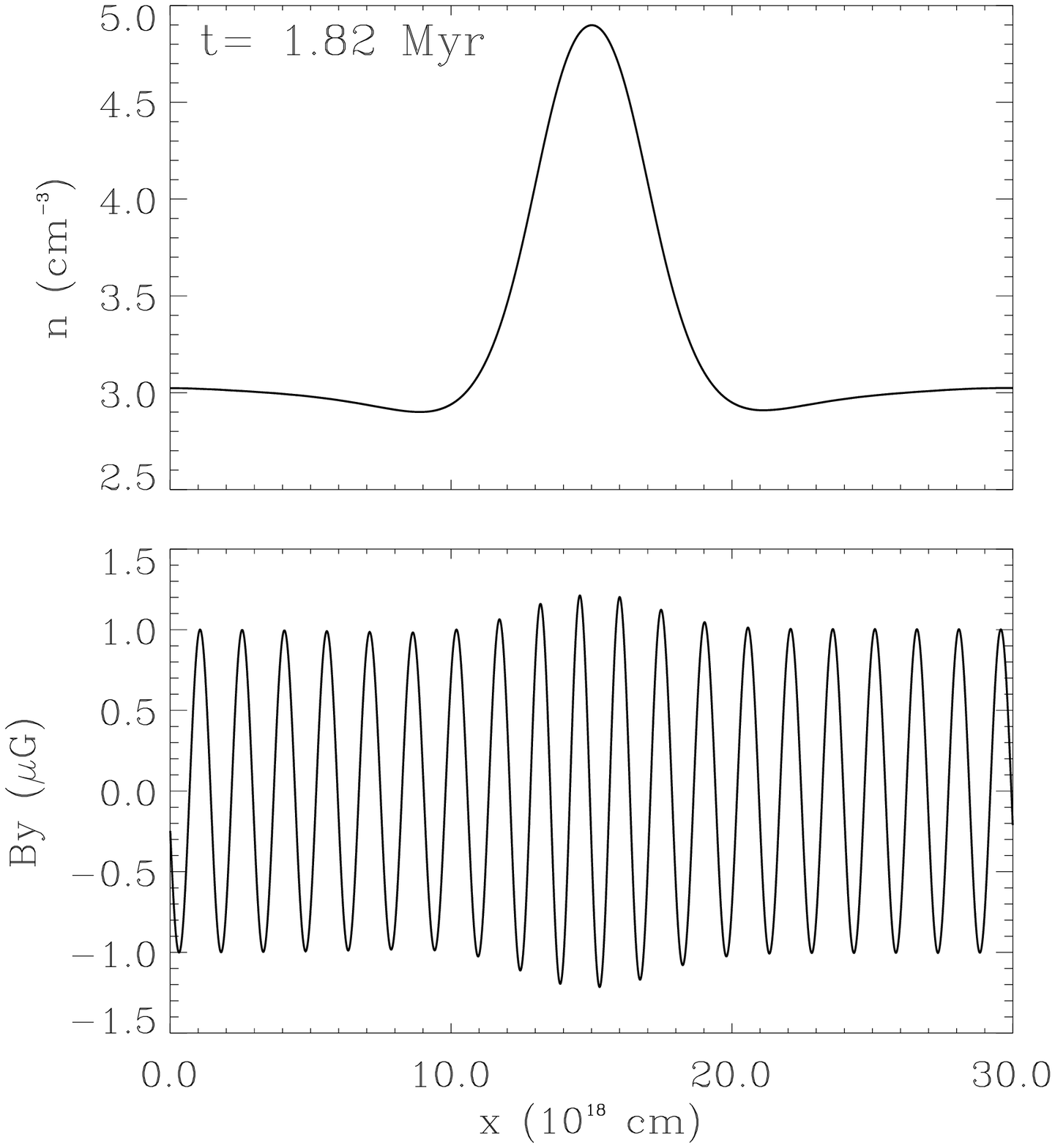}}
\put(0,9){\includegraphics[width=7cm]{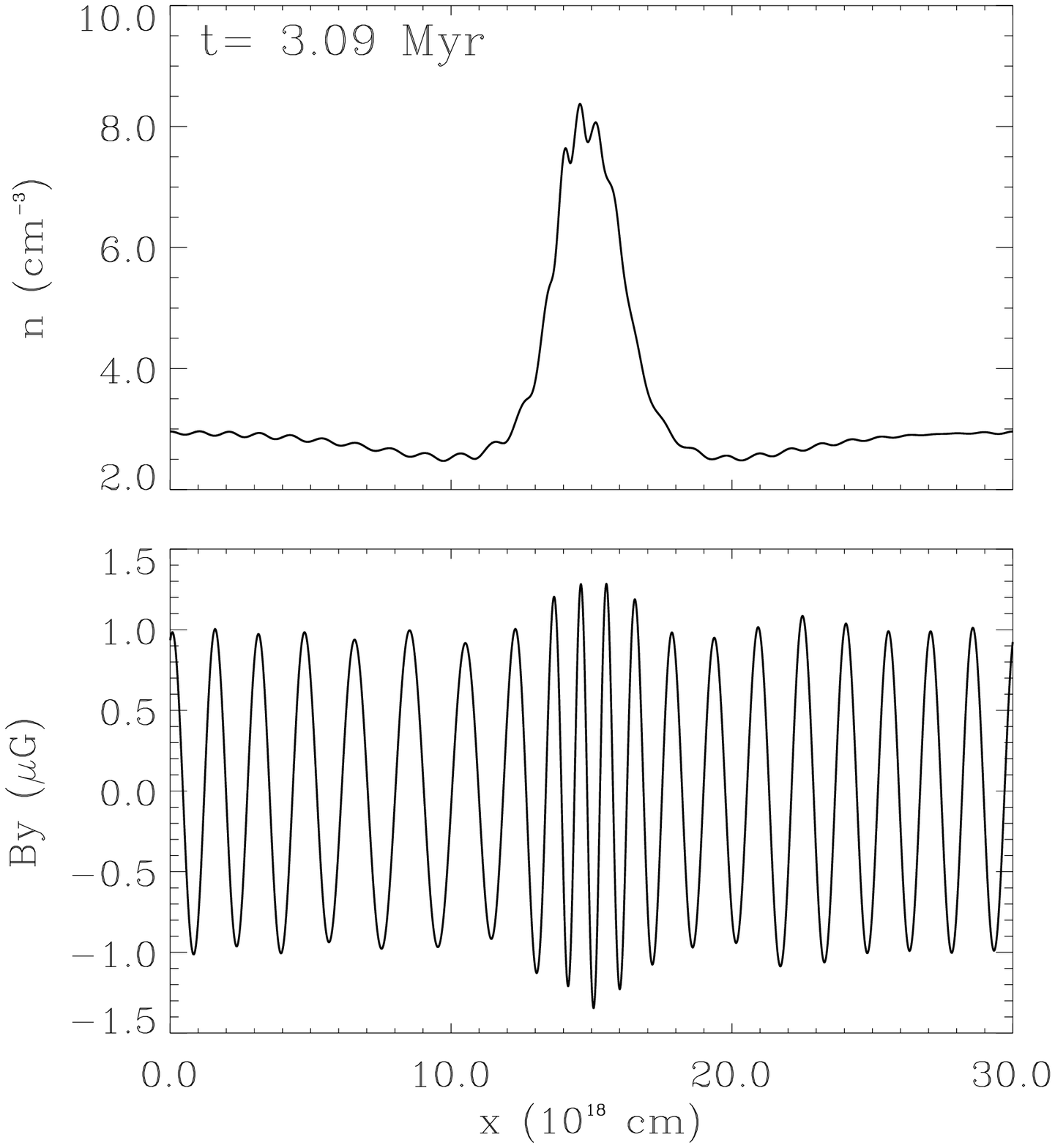}}
\put(0,1){\includegraphics[width=7cm]{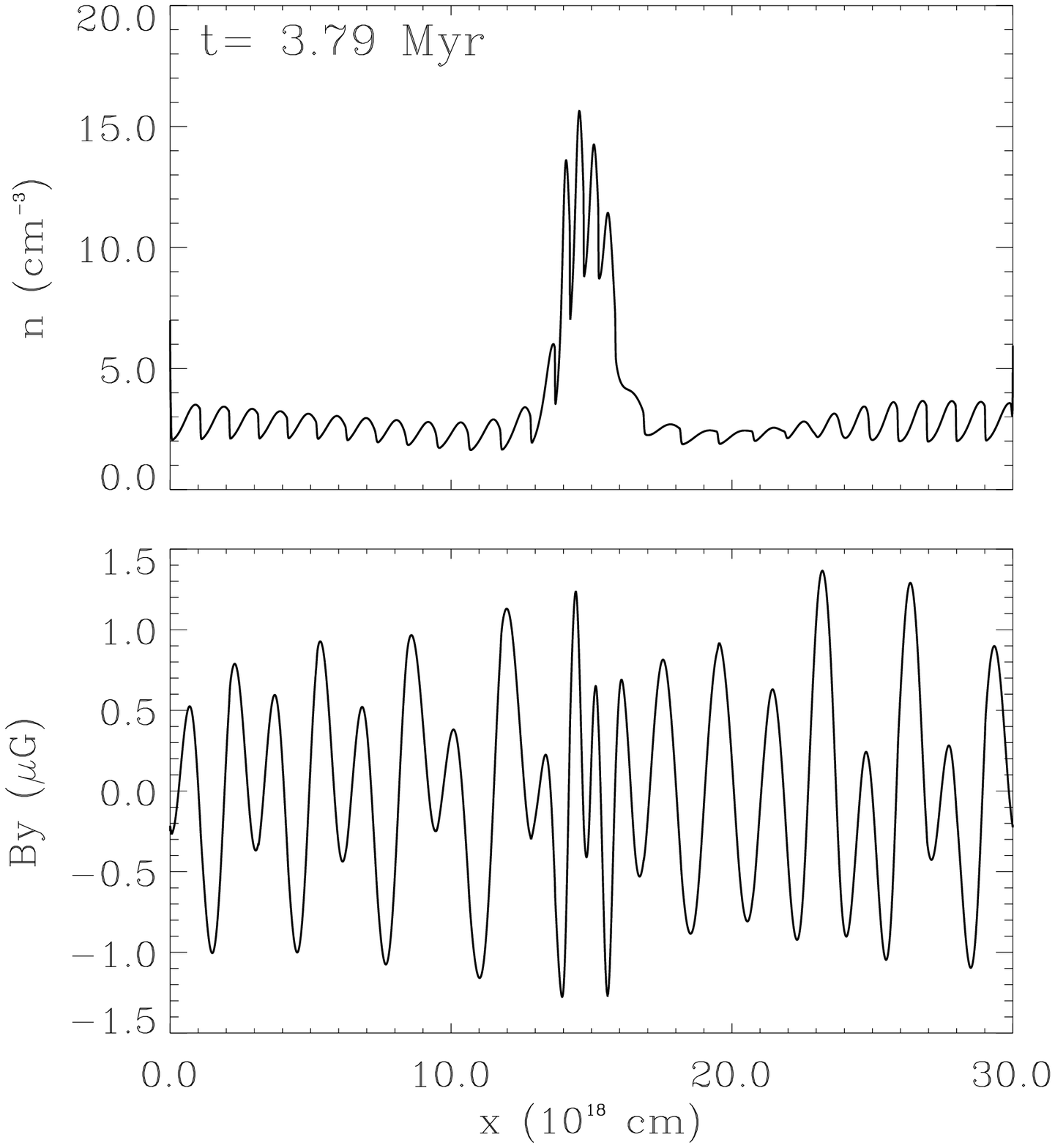}}
\end{picture}
\caption{Density and y-component of the magnetic field for 3 snapshots
illustrating the development of the thermal instability in the
presence of Alfv\'en waves of amplitude $B _\perp=1 \, \mu$G
and a longitudinal magnetic field $B_x=5 \, \mu$G.}
\label{b5a1}
\end{figure}

\begin{figure}
\begin{picture}(0,25)
\put(0,17){\includegraphics[width=7cm]{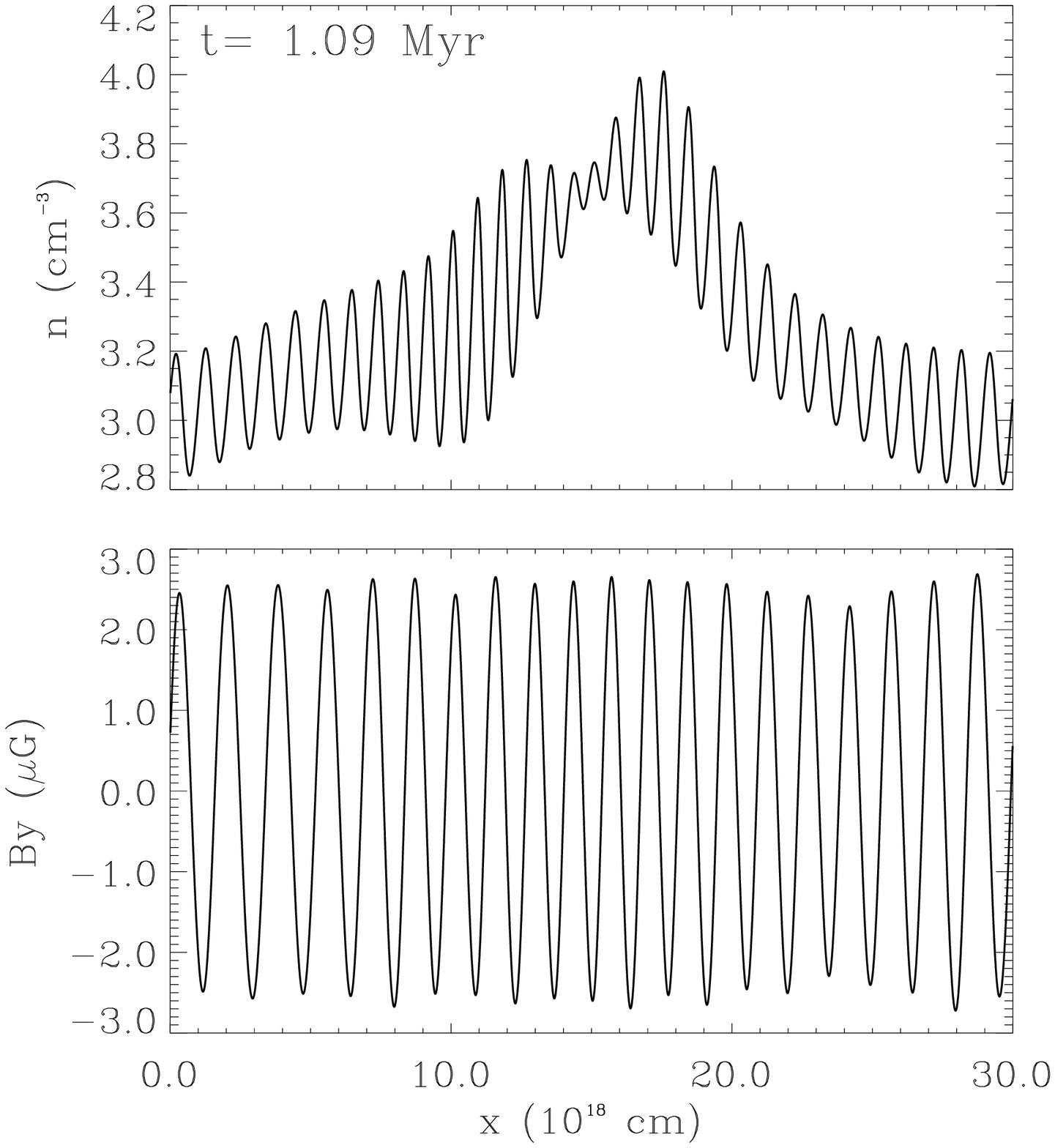}}
\put(0,9){\includegraphics[width=7cm]{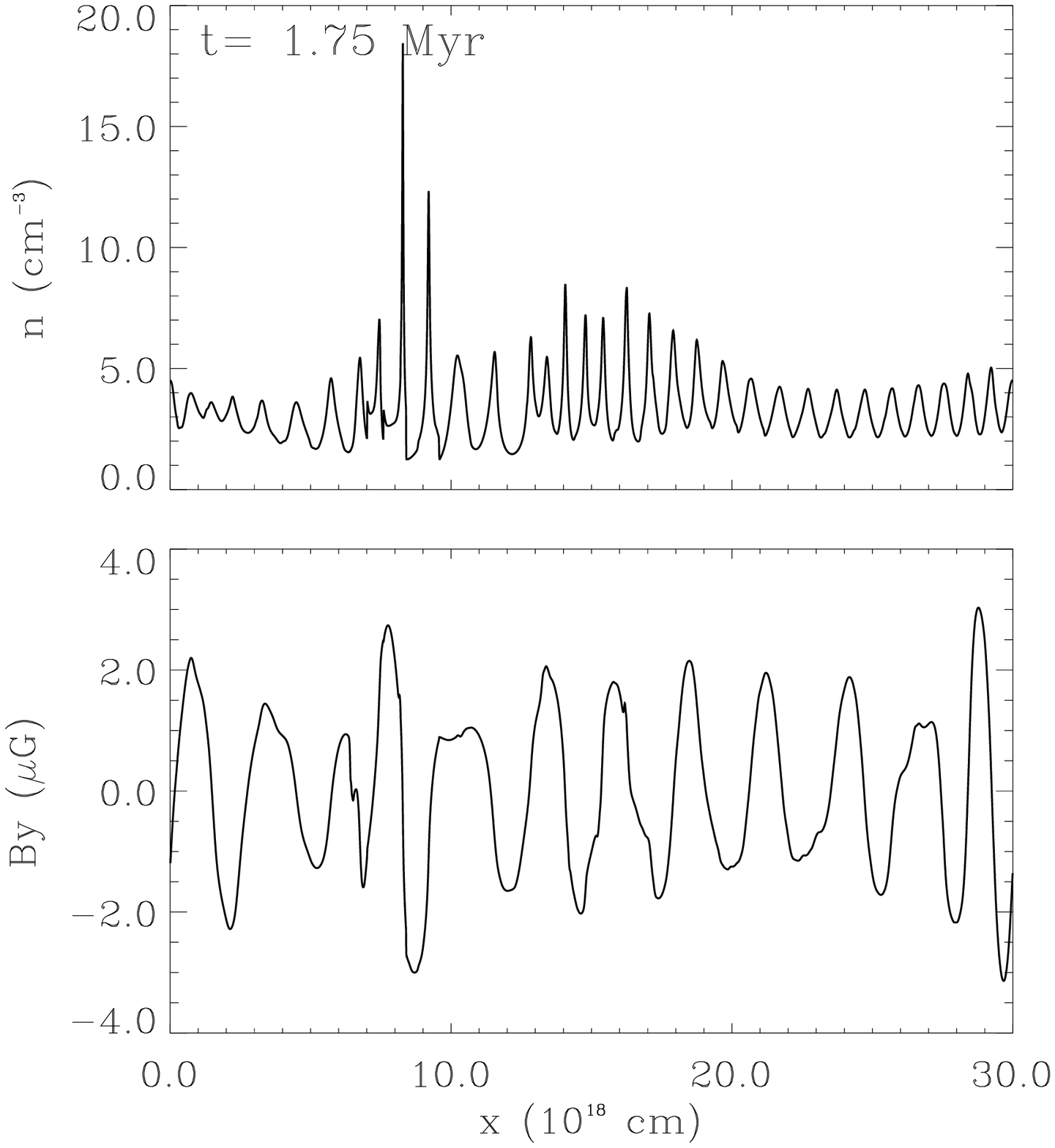}}
\put(0,1){\includegraphics[width=7cm]{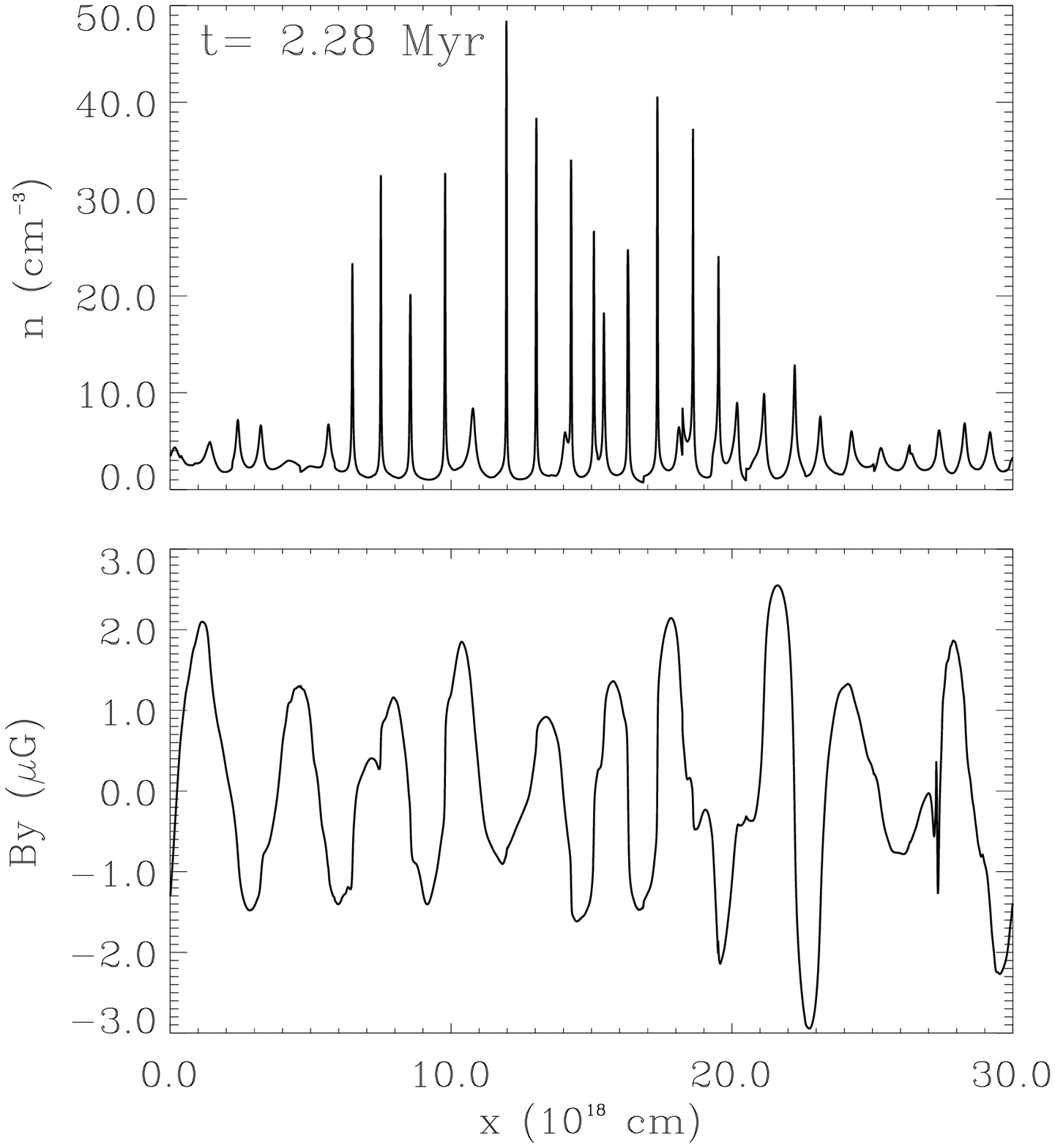}}
\end{picture}
\caption{Same as Fig.~\ref{b5a1} for $B _\perp=2.5 \, \mu$G
and  $B_x=5 \, \mu$G.}
\label{b5a2.5}
\end{figure}

\begin{figure}
\begin{picture}(0,25)
\put(0,17){\includegraphics[width=7cm]{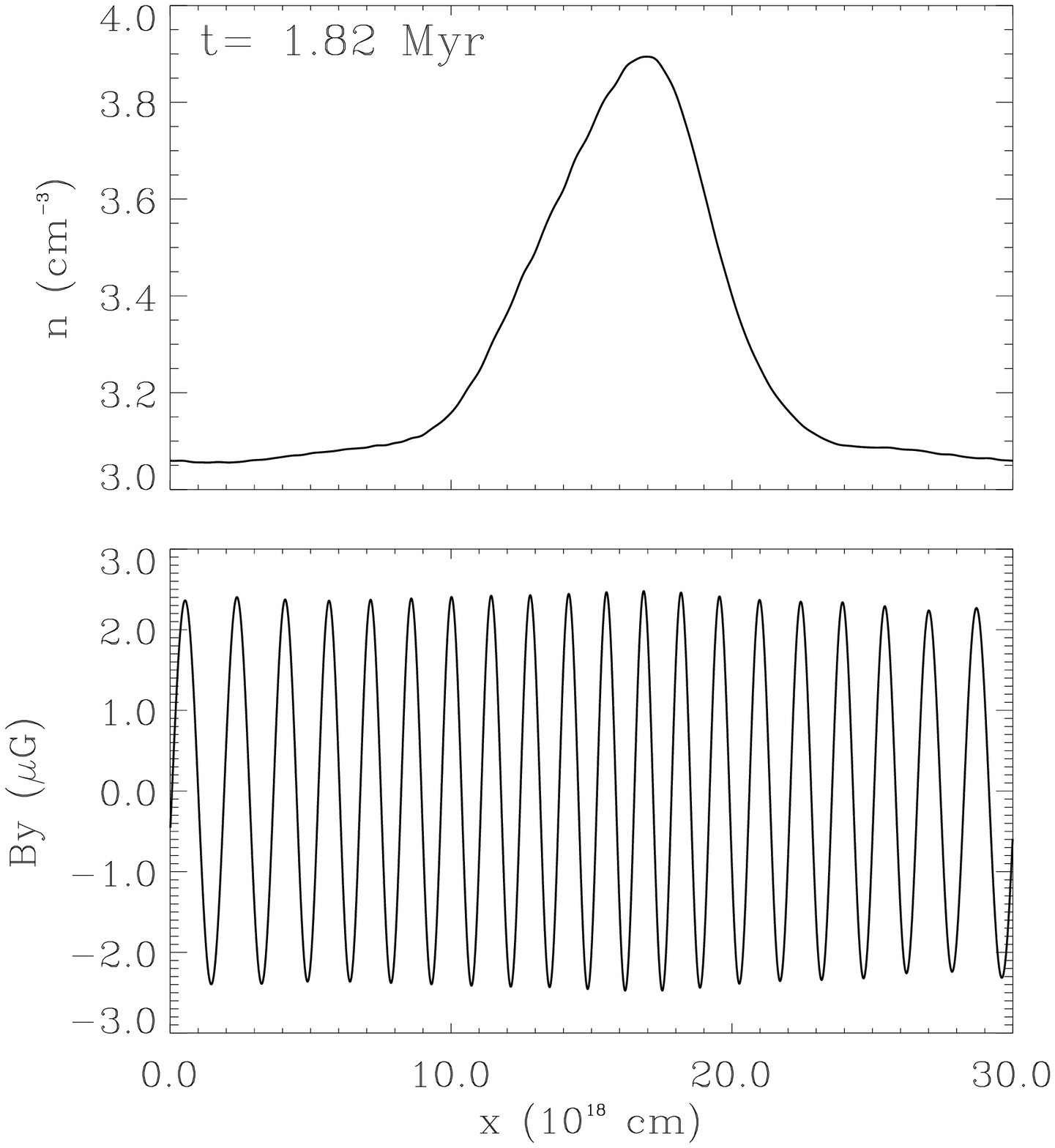}}
\put(0,9){\includegraphics[width=7cm]{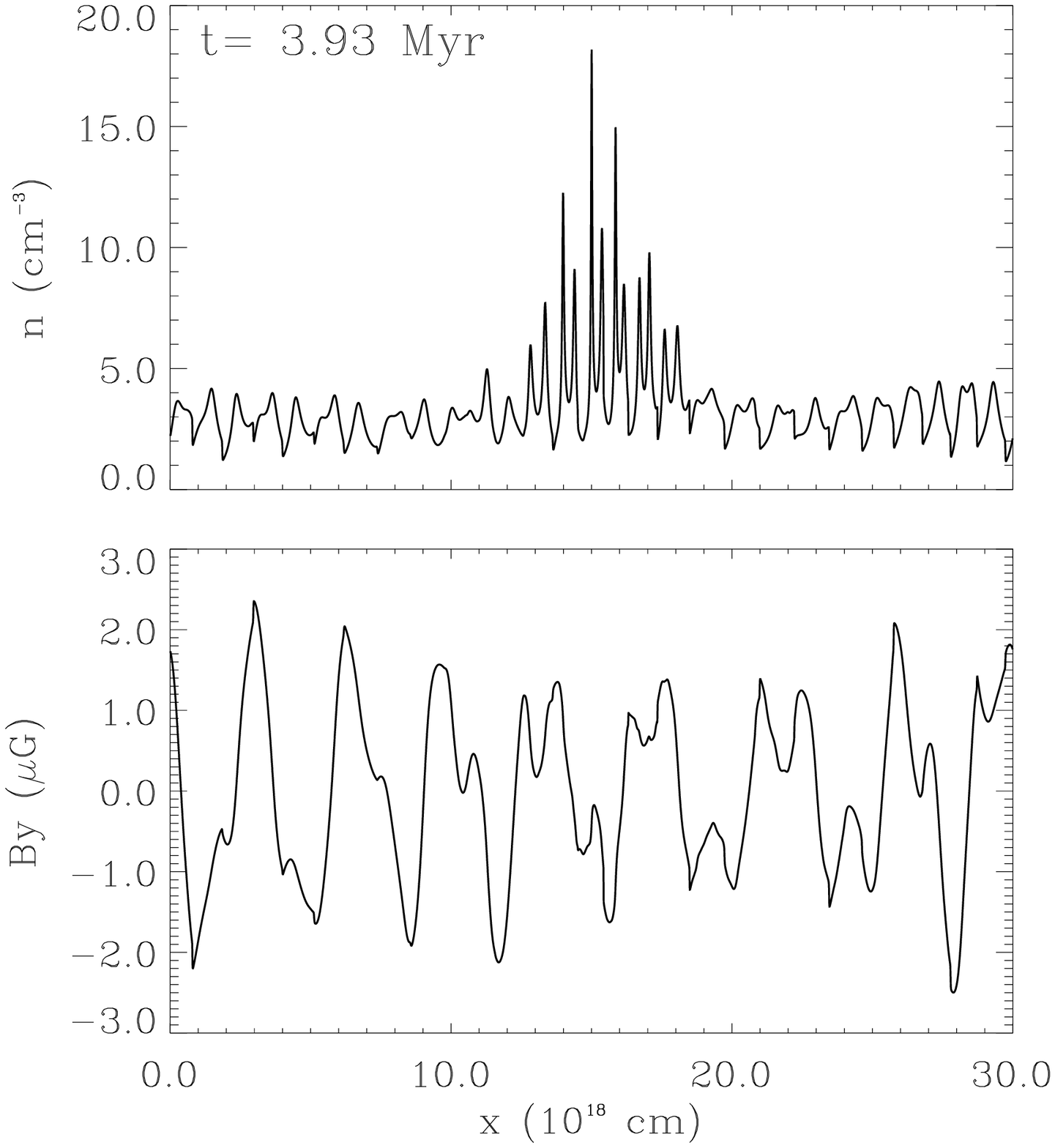}}
\put(0,1){\includegraphics[width=7cm]{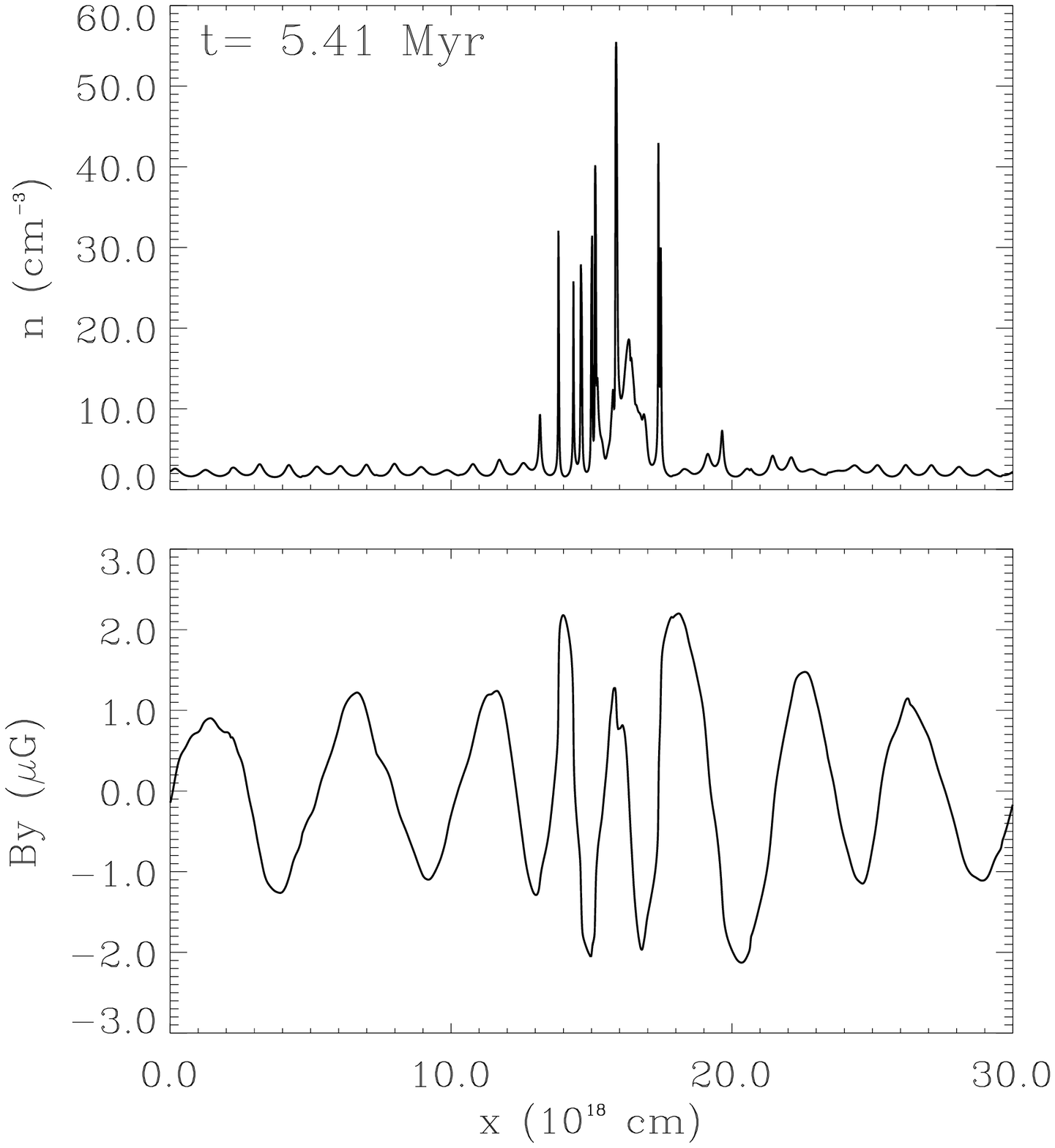}}
\end{picture}
\caption{Same as Fig.~\ref{b5a1} for $B _\perp=2.5 \, \mu$G
and  $B_x=2.5 \, \mu$G.}
\label{b2.5a2.5}
\end{figure}


In order to test the analytic results presented in the previous
section and to investigate the non-linear regime,  
numerical simulations are performed in a slab geometry.

For this purpose we use the 1D adaptive mesh refinement (AMR) code
presented in Hennebelle \& P\'erault (1999, 2000). 
The AMR technique is very helpful to simultaneously resolve  the 
sharp thermal fronts ($\simeq 10^{-3}$ pc) and the
larger scale ($\ge$ 10 pc) involved in the problem.
The code has been extensively tested and the results in the
hydrodynamical case have been closely  compared with high resolution simulations
using a second order  Godunov scheme.
The growth rate for the parametric instability of an Alfv\'en wave
in an adiabatic gas have been calculated and shown to match, within
an accuracy of a few percents, the results of the first panel of
Fig. \ref{growthrate}.

Two numerical experiments are carried out. First, we consider a situation 
for which  the gas is initially thermally unstable and we study the development 
of the thermal instability in the presence of Alfv\'en waves. Although 
these initial conditions are somewhat artificial, they are simple and close 
to the assumption of the analytic analysis, making comparisons 
easier. Second, we setup more realistic 
initial conditions corresponding to a converging flow
 of  thermally stable WNM. In this case the ram pressure of the flow drives
the thermal collapse dynamically.

\subsection{Case of gas initially thermally unstable}
In order to study the effect of the circularly 
polarized Alfv\'en waves on the development 
of the thermal instability, we start the simulation with thermally
unstable gas ($n=3$ cm$^{-3}$ and $T \simeq 500$ K) with density
fluctuations  of amplitude 0.5.
The cooling function is described in Audit \& Hennebelle (2005).
The computational domain, which initially contains 5000 pixels, has
periodic boundary conditions and a length of $3\times 10^{19}$ cm, corresponding
to 20 wavelengths of the initial Alfv\'en wave.

Fig.~\ref{hydro} shows three snapshots of the density field 
in a purely hydrodynamic run (no MHD wave is present).
The initial perturbation grows as a result of the thermal instability. 
At time $t=$2.55 Myr the  density is about three times its initial
value. At time $t=$3.64 Myr a cloud of CNM having a size of about
0.3 pc,  has formed.

Fig.~\ref{b5a1} displays the density and the y-component of the
magnetic field for 3 snapshots showing the development of the thermal
instability in the presence of Alfv\'en waves of amplitude $B _\perp=1 \,
\mu$G and for $B_x=5 \, \mu$G. In this situation the value of
$\omega_c / \omega _0$ is about 0.03, $A=0.2$ and $\beta=0.17$. 
The peak density at time $t=1.82$ Myr is about 4.9 whereas the initial
peak density is 4.5.
In the hydrodynamic case the peak density at the same time is about
6.2 showing that the waves have significantly slowed down  the growth of 
the perturbation, by a factor of about $(6.2-4.5) / (4.9-4.5) \simeq 4$.
Note that interestingly enough, this factor is significantly larger than 
what is predicted by the linear theory which predicts a difference of 
about 10\% (in the case of a perturbation having
an  initial amplitude of $10^{-2}$ we verified that the growth agrees with 
the linear theory). 
After the central density has increased
by a factor of about 2 (panel 2), the waves drastically change the 
structure of the gas and create significant density contrasts ($\simeq$10\%).
The resulting cloud (panel 3) contains density fluctuations of about 
$\simeq$ 50\% its maximum value, and 
is therefore very different from the uniform cloud formed
in the hydrodynamical case. These large fluctuations 
are due to magnetic pressure variations. 
It is worth noting that due to the contraction, 
the waves inside the growing perturbation (panel 2 and 3) have
a larger amplitude and a shorter wavelength than the waves in the
surrounding medium. 
According to the analysis of the preceeding sections,
this effect tends to increase the influence of the waves.

Fig.~\ref{b5a2.5} shows results for $B _\perp=2.5 \, \mu$G and for
 $B_x=5 \, \mu$G. The waves  strongly influence the gas
 evolution. The initial perturbation  is totally 
stabilized by the waves and does not develop (second panel). 
On the contrary, the waves  trigger the formation of  
structures having a wave number $k \simeq k_0 $ (first and second panels). 
These structures keep condensing (panel 3) and finally about 12 CNM clouds 
form. Their size is about 0.03 pc and therefore about 10 times
smaller than the size of the cloud that forms in the hydrodynamical case.
It is worth noting that the formation of these clouds is
significantly (50\%-100\%) faster than in the two previous cases. Therefore in this
range of parameters, the waves destabilize the gas and accelerate the
formation of CNM structures.

Fig.~\ref{b2.5a2.5} shows results for $B _\perp=2.5 \, \mu$G and for
$B_x=2.5 \, \mu$G. In this situation the value of
$\omega_c / \omega _0$ is about 0.06, $A=1$ and $\beta=0.34$.
As can be seen on first panel, the initial perturbation does not grow
(the peak density is smaller at time $t=1.82$ Myr than the initial value).
Unlike the previous case, the Alfv\'en   waves do not efficiently trigger  
the formation of structures at $k \simeq k_0$ (see first panel).
However since the waves are unable to stabilize small wavelengths,
the initial perturbation  breaks down in several structures which
finally develop (panel 2). This nevertheless occurs  after a
significant delay showing that in this situation, the gas spent a
longer amount of time in the  thermally  unstable state. At time
$t=5.41$ Myr  about 6 small structures of size $\simeq 0.03$ pc 
 and one larger structure of size $\simeq 0.3$ pc have  formed.

\begin{figure}
\includegraphics[width=7cm]{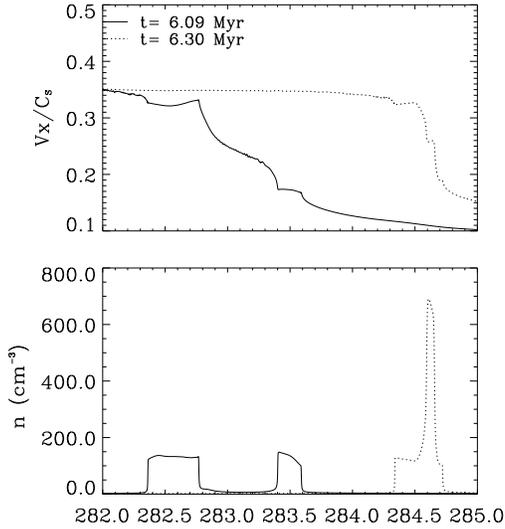}
\caption{ Spatial zoom showing the thermal condensations which have been 
induced by the large scale converging flow in the hydrodynamical case. Two snapshots are displayed.}
\label{converging_hy}
\end{figure}

\begin{figure}
\begin{picture}(0,20.5)
\put(0,10.5){\includegraphics[width=7cm]{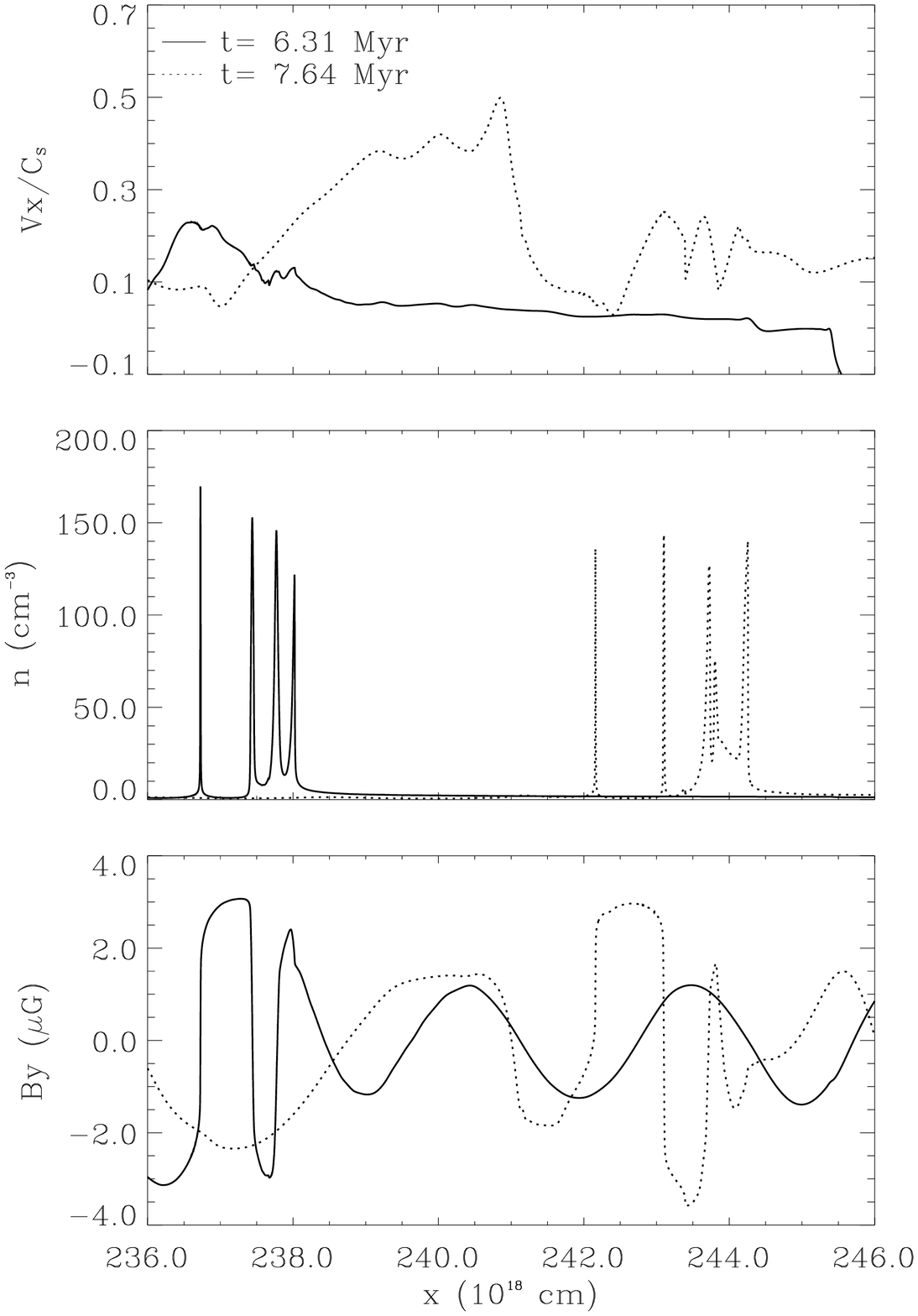}}
\put(0,0.5) {\includegraphics[width=7cm]{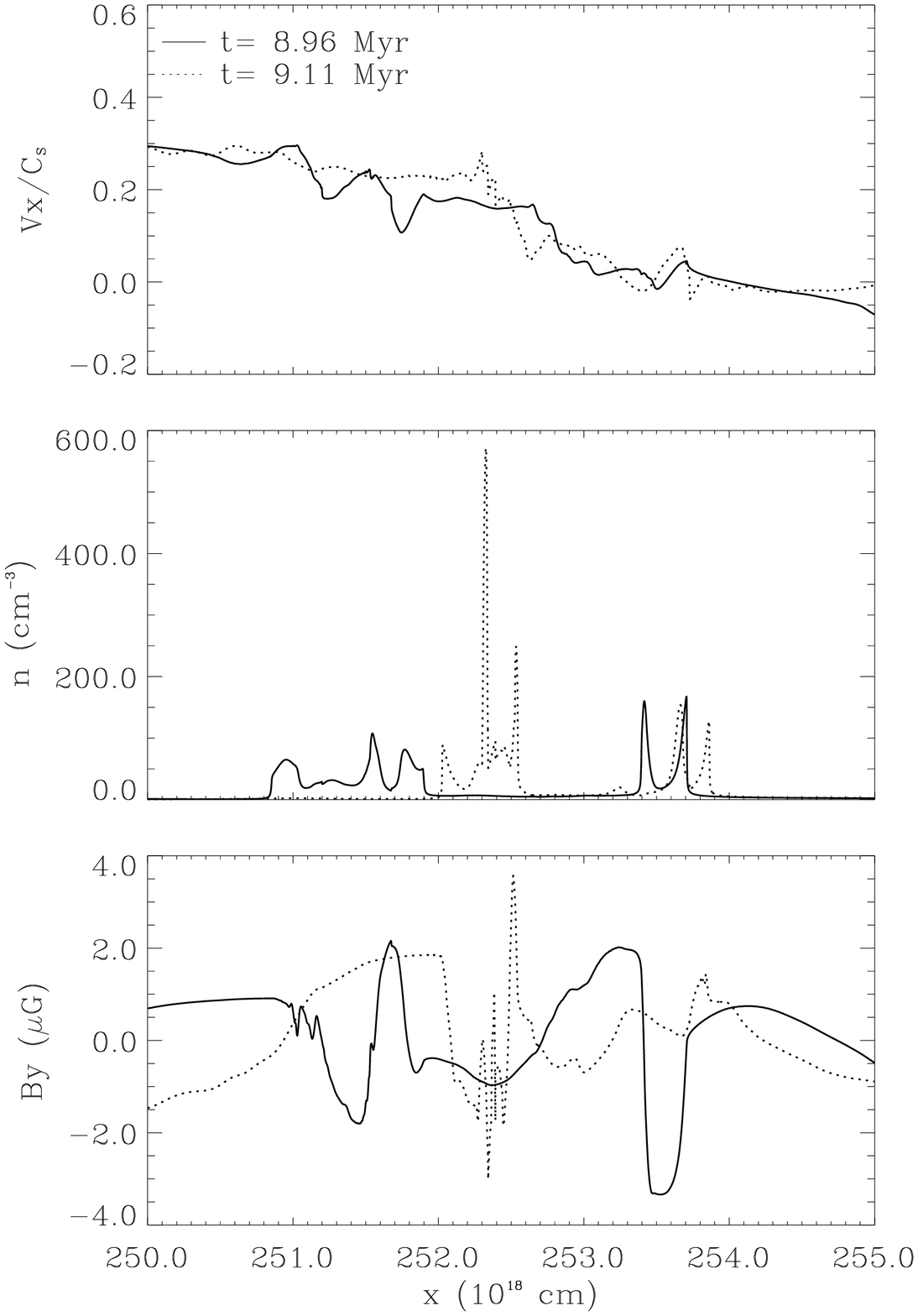}}
\end{picture}
\caption{
Same as Fig.~\ref{converging_hy} in the presence of 
circularly polarized Alfv\'en waves  (see text for detail).
Four snapshots are displayed.}
\label{converging}
\end{figure}

\subsection{Case of thermal condensation dynamically induced}
We now consider a converging flow of WNM (Hennebelle \& P\'erault
1999) in a simulation box of length 150 pc. The peak velocities are  
2.2 $C_{wnm}$ and -2 $C_{wnm}$ and the peak to peak distance is about 60 pc.
Two simulations are performed. The first one is purely hydrodynamical,
while the second one starts with an Alfv\'en waves of amplitude 2.5
$\mu$G with 100 spatial periods in the integration domain.  The total magnetic 
intensity is 5 $\mu$G  and the longitudinal one $B_x \simeq 4 \mu$G.
An initial resolution of 25,000 pixels is used in order to 
ensure an accurate description of the Alfv\'en wave (it corresponds to 250
pixels per period). 

Fig.\ref{converging_hy} shows the hydrodynamical result. At time
$t=6.09$ Myr, two clouds having a size of about 0.1 pc have  
formed. They present weak density gradients and have no significant
internal velocity. The two clouds  
have a relative velocity of about 0.15 km/s and  undergo 
a collision at time $t=6.3$ Myr. A shock-compressed layer forms with
a density of about $\simeq 700$ cm$^{-3}$ and a  length $\simeq 0.02$
pc. During the time of the collision,  the structure presents a
stiff velocity gradient.

Fig.\ref{converging} displays four snapshots of the longitudinal
velocity field, the density and  the y-component of the magnetic
field  in the MHD case. 
The results displayed in the first panel confirm the trends observed in the 
numerical experiments of Sect.~3.1.  The gas  fragments into few
small CNM structures  having a physical length as small as few 0.01
pc (note that due to the AMR scheme, these structures are well described). 
This situation is very different from the hydrodynamical case for
which the structures  are much larger and uniform.
A comparison between the 2 times displayed in first panel of 
Fig.~ \ref{converging} reveals that
all the structures do not form at the same time. The structure of the magnetic
field deserves attention. It varies very rapidly in the new born condensations
($x=242$ cm and $x=243$ cm), therefore compressing them. 
It varies less stiffly near older structures  ($x \simeq 244$ cm) since the 
field lines had time to unbend. Between the 
clouds,  the magnetic field is much more uniform. This is a consequence 
of the fact that the Alfv\'en speed is about 10 times larger in the WNM than
in the CNM. It is also clear that the intercloud magnetic pressure
plays an important role in preventing the merging of the CNM structures
therefore maintaining the complexity of the flow and organizing it into groups
of structures rather than into a single cloud.
In these groups of structures the longitudinal velocity dispersion
is not negligible, unlike in the hydrodynamical case. It 
is about 0.1-0.2 km/s, i.e. few times the sound speed or the Alfv\'en speed 
of the CNM. This velocity dispersion is due to the transfer of  magnetic energy 
into longitudinal motions because of the magnetic pressure fluctuations.
We stress the fact that observing such group of structures with a low spatial 
resolution ($\simeq $0.2 pc) may lead to a rather different picture, 
namely a broad, uniform and turbulent (having a Mach number M$\simeq 1-2$)
CNM structure.

The second panel of Fig.\ref{converging} shows two later snapshots of the same
numerical experiment. The group of structures which is seen in the first panel
is now located at $x \simeq 251.5$ pc. Large fluctuations ($\simeq 100$\%)
of density  and magnetic field are still present as well as a 
longitudinal velocity dispersion of about $0.1-0.2$ km/s.
Another  smaller group  of structures  ($x \simeq$ 253.5 pc) has formed.
At time $t=9.11$ Myr (dotted line), the first cloud undergoes a large 
density fluctuation ($n _{\rm max} \simeq$ 600 cm$^{-3}$) at a
scale of about 0.01 pc. At the same time the cloud is compressed, because of 
the (magnetic) interaction with the other cloud, so that its length is divided 
by a factor $\simeq 2$. An interpretation based on the
previous analytic results is that the amplitude and wavenumber of
the  Alfv\'en wave in the cloud increasing as a result of the
compression, the decay instability is triggered leading to 
a larger density fluctuations.  Interestingly, in contrast with the 
stiff and large fluctuations undergone by $B_y$,  the longitudinal 
velocity field remains relatively smooth. 

\section{Discussion and conclusion}
The analysis presented in Sect.\ref{sec:num} as well as the numerical 
experiments discussed in the previous section show that even modest
amplitude Alfv\'en waves
may have a strong impact on the structure of the 
multiphase ISM. This study focused on circularly polarized parallel
propagating Alfv\'en waves, mainly to allow analytic calculations.
These waves, which are exact solutions of
the MHD equations, are very weakly dissipative and are therefore very
likely to be present in the ISM. 

Their  effects depend on $\beta$, 
$\widetilde{\omega_c}$ and $A$, respectively the square ratio of the
sound to the Alfv\'en speeds, the ratio between the wave temporal period and 
the cooling time and the wave amplitude. Depending on the 
values of these parameters, these waves may:
 i) stabilize the wavelengths larger than that of the  Alfv\'en
 wave, $\lambda _{\rm AW}$, that   
 would otherwise be thermally unstable, therefore enhancing the fraction 
of thermally unstable gas in the ISM , ii)  destabilize the wavelengths
comparable to $\lambda _{\rm AW}$ and thus fragment the CNM into several
spatially correlated  small clouds ,  iii) induce strong density
fluctuations within preexisting  CNM structures (up to 10 times the mean density),
iv) maintain an Alfv\'enic velocity dispersion
within the CNM structures by pumping their energy into longitudinal motions.
Finally magnetic pressure tends to prevent the merging between CNM
clouds.

These effects are not observed in one-dimensional purely hydrodynamical
simulations leading to larger and almost uniform structures with
weak internal motions. In two dimensions (Koyama \& Inutsuka 2002, 
Audit \& Hennebelle 2005) the situation is more complex
due to the role of turbulence, with coexistence of 
thermally unstable gas and small-scale structures, but the latter are
still relatively uniform and do not contain significant
velocity dispersion. According to the present analysis, MHD waves may  
enhance the thermal fragmentation found in these simulations, 
maintain an Alfv\'enic velocity dispersion and generate large
density fluctuations within the CNM structures.

Recent observational progress has revealed the presence of interesting features
and of a large quantity of thermally unstable gas 
(e.g.  Heiles 2001, Heiles \& Troland 2003,  Miville-Desch\^enes et
al. 2003).  Heiles (1997) summarizes
and discusses observations of tiny small-scale structures and more
recently Braun \& Kanekar (2004, 2005) and Staminorovic 
\& Heiles (2005) report the detection of very low column density CNM clouds.

Qualitatively, the present study as well as the 2D simulations of
Koyama \& Inutsuka (2002) and Audit \& Hennebelle (2005) show
similar features including thermally unstable gas (also observed by
Gazol et al. 2001 in simulations at larger scale), 
low column density structures (down to and even smaller than
 $10^{18}$  cm$^{-2}$) and large density fluctuations at small scale
(up to 10$^3$ cm$^{-3}$). 
Small-scale structures and large density fluctuations appear
to be a natural outcome of the two-phase nature of the flow. 
In the absence of turbulence or magnetic field, CNM structures have
a typical  length of about 0.1 pc  (or a column density of $\simeq 
10^{19}$ cm$^{-2}$) (Audit \& Hennebelle 2005). In the presence of
waves or turbulence, a growing structure can however 
fragment into smaller pieces. Moreover, since both the sound
and the Alfv\'en speed change by a factor $\simeq$ 10 
between the WNM and the CNM, with a transition occurring on a short
distance of the order of a Field length, supersonic or
superalfv\'enic CNM motions are expected to be more frequent.  

No systematic and quantitative study attempting to closely compare
simulations and observations has been carried out yet.  
This is clearly a major challenge for the future. 

\begin{acknowledgements}
This work has received partial financial support from the French
national program PCMI.
\end{acknowledgements}

\end{document}